\documentclass[reqno]{amsart}

\usepackage[margin=3cm]{geometry} %set the margins

\usepackage{bm}
\usepackage{bbm} %For blackboard fonts with numerals. To define the identity function \mathbbm{1}
\usepackage{hyperref} %[hidelinks] hides the fluorescent boxes and makes all colors black
\hypersetup{
    colorlinks = true,
    linkcolor={red!50!black},
    citecolor={blue!50!black},
    urlcolor={blue!50!black}
}
\usepackage{cite}

\usepackage{siunitx}
\usepackage{mathtools}
\usepackage{amssymb}
\usepackage{mathrsfs}
\usepackage[normalem]{ulem}
\usepackage{enumitem}
\usepackage{amsthm} %For defining `Remarks, Definitions, Theorems, Lemmas and Corollaries'
\usepackage[capitalise]{cleveref}

\usepackage[dvipsnames]{xcolor}

\usepackage{xparse} %allows more powerful and customisable \newcommand called \NewDocumentCommand

\newcommand{\be}{\begin{equation}}
\newcommand{\ee}{\end{equation}}
\newcommand{\e}[1]{\label{#1}\end{equation}}
\newcommand{\bea}{\begin{eqnarray}}
\newcommand{\eea}{\end{eqnarray}}
\newcommand{\ea}[1]{\label{#1}\end{eqnarray}}
\newcommand{\bse}{\begin{subequations}}
\newcommand{\ese}{\end{subequations}}

\newcommand{\px}{\ensuremath{\func{p}{x;\theta}}}

\newcommand{\prior}[1][\theta]{\ensuremath{\func{\pi}{#1}}}

\newcommand{\emse}[1]{\ensuremath{\func{\overline{mse}}{#1}}}

\newcommand{\iu}{i\mkern1mu}
\newcommand{\func}[2]{\operatorname{#1}\left[#2\right]} %Defining an arbitrary function that works as a latex operator \sin etc and has square brackets for the argument
\newcommand{\smashfunc}[2]{\operatorname{#1}\!\left[#2\right]} %Defining an arbitrary function that works as a latex operator \sin etc and has square brackets for the argument
\newcommand{\funcwrt}[3]{\operatorname{#1}_{#2}\left[#3\right]} %Defining an arb. function as a latex operator \sin etc and has square brackets for the argument
\newcommand{\funcup}[3]{\operatorname{#1}^{#2}\!\left[#3\right]} %Defining an arb. function as a latex operator \sin etc and has square brackets for the argument
\newcommand{\setfunc}[2]{\left\{\operatorname{#1}\left[#2\right]\right\}} %Defining an arb. function on a set that works as a latex operator \sin with square brackets for the argument
\newcommand{\funclims}[3]{\operatorname*{#1}\limits_{#2}\left[#3\right]} %Defining an arbitrary function that works with limits \sin and has square brackets for the argument
\newcommand{\ket}[1]{\left| #1 \right>}
\newcommand{\bra}[1]{\left< #1 \right|}
\newcommand{\braket}[2]{\langle#1|#2\rangle}

\newcommand{\evalwrt}[2]{\mathbb{E}_{#1}\left[#2\right]} %Expected value
\newcommand{\est}[1]{\hat{#1}} %Definition of an estimator

\newcommand{\antiprob}[1]{\smashfunc{\pi^{\circ}}{#1}}
\newcommand{\antidist}[1]{\smashfunc{d^{\circ}}{#1}}

\newcommand{\Op}[1]{\bm{#1}} %Definition of operator notation

\newcommand{\comp}[1]{#1^{\complement}}

\newcommand{\extnneg}{[0,+\infty]} %The extended non-negative real numbers
\newcommand{\dist}[1]{\func{d}{#1}} %Shortcut for distance function
\newcommand{\MS}[1][M]{\ensuremath{(#1,\mathrm{d})}} %Shortcut for metric space
\newcommand{\Md}[1][M]{\ensuremath{#1_{\mathrm{d}}}} %Concise form of a metric space
\newcommand{\norm}[1]{\lVert #1 \rVert} %Definition of metric diameter or norm
\NewDocumentCommand{\setvalO}{ O{O} O{I} }{\setfunc{\text{\ensuremath{#1}} \mapsto Val}{#2}} %Shortcut for set of values with origin

\newcommand{\MB}[1][\mathcal{B}_x]{\ensuremath{(X,#1)}} %Shortcut for measurable space on a set #1, with sigma algebra \mathcal{B}
\newcommand{\PS}[1][\mathcal{B}_x]{\ensuremath{(X,#1,\func{p}{\cdot})}} %Shortcut for probability space on a set X, with sigma algebra #1, and probability measure p.
\NewDocumentCommand{\ID}{ O{I} o } %Shortcut for metric interval, default is (I, {Val[I]})
{ \IfNoValueTF {#2}
{\ensuremath{(#1,\setfunc{Val}{#1})}}
{\ensuremath{(#1,\setfunc{Val}{#2})}}
}
\usepackage{titlesec} %titlesec package
\titleformat{\section}{\centering\Large\sffamily}{\thesection.\quad}{0em}{}{}
\titlespacing*{\section}{0pt}{*2}{*1}
\titleformat{\subsection}{\raggedright\large\bf\color{blue!30!black}}{\thesubsection.\quad}{0em}{}{}
\titlespacing*{\subsection}{0pt}{*1}{*0.5}
\titleformat{\subsubsection}{\raggedright\normalfont\bf}{\thesubsubsection.\quad}{0em}{}{}
\titlespacing*{\subsubsection}{0pt}{*1}{*0.5}

\newtheorem{theorem}{Theorem}

\newcommand{\thistheoremname}{}
\newtheorem{genericthm}[theorem]{\thistheoremname}

\theoremstyle{definition}
\newtheorem{definition}{Definition}
\newtheorem{example}{Example}
\newtheorem{remark}{Remark}
\newtheorem{assumption}{Assumption}

\newtheorem*{thesis}{Church-Turing Thesis}

\crefname{assumption}{Assumption}{Assumptions}

\NewDocumentEnvironment{genthm}{m m}{%
    \begin{#1}[#2]%
}{%
    \end{#1}%
}

\makeatletter
\newcommand{\myitem}[1]{%
  \refstepcounter{enumi}%
  \item[#1]%
  \edef\@currentlabel{#1}%
}
\makeatother

\makeatletter
\newcommand{\grayub}{}
\def\grayub#1{%
  \@ifnextchar_%
    {\@grayub{#1}}
    {\@latex@warning{Missing argument for \string\grayub}\@grayub{#1}_{}}%
}
\def\@grayub#1_#2{%
    \colorlet{currentcolor}{.}%
    \color{gray}%
    \underbrace{\color{currentcolor}#1}_{\color{gray}#2}%
    \color{currentcolor}%
}
\makeatother

\begin{document}
\title{Genuine certifiable randomness from a black-box}
\maketitle

\begin{centering}
\author{Liam P. McGuinness}\\
\address{Institute for Quantum Optics, Ulm University, 89081, Ulm, Germany}\\
\email{Email: \href{mailto:liam@grtoet.com}{liam@grtoet.com}}\\
\end{centering}

\begin{abstract}
Randomness is intrinsic to quantum mechanics; the outcome of a measurement on a quantum state is a random variable. This feature has been applied to randomness certification, where one party must decide whether the data they receive is truly random \cite{Pironio2010, Acin2016, Liu2025}. However existing demonstrations are not black-box, to avoid falsely certifying deterministic data assumptions must be made on how the data was generated. Here we demonstrate genuine randomness certification in the black-box setting -- one in which no deterministic adversary, even with unlimited computational power, will succeed in getting their data certified. We use it to provably generate random numbers using only measurements on single particle states and without a random seed.%with several benefits over current approaches. We certify randomness using only measurements on single particle states and without a random seed whereas current approaches use measurements on entangled states and require an initial source of randomness.
\end{abstract}

Randomness certification is the task of verifying that some data is truly random and not the output of a deterministic, possibly pseudorandom function. One party -- the \emph{prover} -- produces data and a second party -- the \emph{verifier} -- must decide whether or not the data was randomly produced. More specifically, in \emph{black-box} randomness certification, the verifier makes their decision without `looking at' the prover or making any assumptions on their physical state, i.e. treating the prover as a black-box. Genuine black-box randomness certification is thought to be impossible; the reasoning being that any randomly generated finite data string can also be produced by a deterministic machine. This logic, however, is flawed. 

Clearly, in order to prove the data is truly random, the prover must \emph{do something} with randomness that cannot be done deterministically and the verifier must be able to distinguish between the two. A first step in deciding the proposition therefore, is to define how randomness enters the model, since without this element the point is moot. We use the model of a \hyperlink{Quantum Randomized Turing Machine}{quantum randomized Turing machine}; a computing device operating according the postulates of quantum mechanics. Quantum randomized Turing machines contain deterministic Turing machines as a subset but incorporate randomness through the \hyperlink{Born rule}{Born rule} -- the outcome of a measurement on a quantum state is a random variable. Indeed, recent work using Bell inequalities \cite{Pironio2010, Acin2016} or quantum computers \cite{Brakerski2021, Mahadev2022, Aaronson2023, Liu2025} has taken this approach to ostensibly certify randomness. But these protocols are not black-box; additional physical and computational constraints beyond the model must be placed on the prover before the verifier can certify the data as truly random. In all that follows, randomness certification is assumed to be black-box unless otherwise stated.

To get a feel for why randomness certification is difficult, even in a quantum mechanical model of computation, think of the prover as a computer with the set of computable functions denoting all the things they can do. Distinguishing randomness from determinism requires the existence of a function that is probabilistically computable with bounded error on a (quantum) randomized Turing machine but not computable to the same error on a deterministic Turing machine. Now such a function, if it existed, would overturn the Church-Turing thesis. % If such a function existed it would upheave a cornerstone computer science. In direct contrast, the 
\begin{thesis}
Any function computable by a physical machine (including quantum machines) can be computed on a deterministic Turing machine to arbitrary precision\ldots although perhaps not efficiently.%\footnote{As long as we don't require efficiency the Church-Turing thesis remains the same when replacing a deterministic Turing machine with a probabilistic Turing machine.}.
\end{thesis}
\noindent Yet despite the minor obstacle posed by the Church-Turing thesis, the idea of using quantum mechanics to certify that some data was randomly generated contains a key insight. \emph{Ultimately, two ingredients are needed for randomness certification:}
\begin{itemize}
\item A device for generating randomness.
\item A means of forcing the prover to use said device.
\end{itemize}
The Born rule provides the first ingredient, whereas we introduce the second by way of analogy. In keeping to the box theme, let the source of randomness be a box that outputs random samples from a probability distribution. Instead of sending the prover the input to a challenge function and asking them to calculate the output, as is typical in verifier-prover interactions, the verifier can send the prover a single-use random box and ask for a sample from the probability distribution. If the `exterior' of the box provides no information on the probability distribution and it really is single-use, then the only option for the prover is to sample from the ouput of the box and return that sample to the verifier. The perspective taken here is that this analogy fully captures both deterministic and probabilistic models of computability. We further stress that the use of quantum mechanics here does not place physical limitations on a classical computer, rather it provides a coherent mathematical framework that extends deterministic computation to probabilistic computation. Allowing functions to have random inputs generalizes deterministic functions to inputs with non-trivial probability distributions. In particular, deterministic function evaluation corresponds to the special case of a randomized Turing machine \emph{reading-in} a non-random input string. 

Much of the paper is devoted to making this analogy rigorous. As a preview, in quantum mechanics i) individual \hyperlink{non-separable}{non-separable} quantum states are the equivalent of single-use random boxes, ii) measurements on quantum states output samples from a probability distribution and iii) certain measurements on quantum states yield deterministic outputs. %\footnote{In particular rotations of a quantum state (the random box) -- called unitary transformations in quantum theory -- modify the probability distribution.}.
We use the machinery of statistical parameter estimation, framing randomness certification not as a computational task but as an estimation problem. The prover is sent a quantum state -- a random box -- and asked to estimate some parameter $\theta$ of this quantum state. %Think of this as the verifier sending an unknown signal to the prover, which they must estimate. 
Although this approach is foreign to complexity theory, therein lies its value. Using a technique outside of standard complexity analysis we prove the following theorem: with measurement data -- a sample from the box -- the prover's estimate can lie in a region that cannot be accessed without a measurement, regardless of the prover's computational power\footnote{Since the outcome of a quantum measurement is the only random component in this model, the terms ``no measurement'' and ``no randomness'' are equivalent and can be freely interchanged. However, the converse does not immediately hold, some measurements are not random.}.

%The verifier can use this method to create the input of their function. Or send it in a box to the prover. The only way the prover can open the box, or use the input is to generate a random number. Like a trap-function.

\subsection*{The Born rule and randomness certification}
Randomness is a fundamental feature of \emph{nearly} all quantum measurements. It arises from the \hyperlink{Born rule}{Born rule}: given a quantum state $\ket{\psi}$ in a complex Hilbert space $\mathscr{H}$, the probability to measure an outcome $x$ associated with a Hermitian operator $\Op{X}_x$ is
\be\label{eq:Born} \func{p_{\ket{\psi}}\!}{x} := \bra{\psi}\Op{X}_x^{\dagger}\Op{X}_x\ket{\psi}.\ee
If $\ket{\psi}$ is not an eigenstate of any measurement operator $\Op{X}_x \in \{\Op{X}_x\}$, then, by the postulates of quantum mechanics, the measurement outcome must be random. That is, $x$ is a random variable with probability distribution such that $\func{p_{\ket{\psi}}\!}{x} < 1, \forall x \in X$, and $\sum_{x \in X} \func{p_{\ket{\psi}}\!}{x} = 1$.

In practice, the Born rule has the following implication for randomness certification. If the verifier is convinced the data they receive is the outcome of a quantum measurement and the measurement was in some sense non-trivial, then they can certify the data as random. More formally, assuming that quantum mechanics is correct, randomness certification reduces to proving two statements:
\begin{enumerate}
\myitem{(M1)}\label{item:M1} The prover performed a measurement $\{\Op{X}_x\}$ on $\ket{\psi}$; the measurement outcome $x$ serves as a witness.
\myitem{(M2)}\label{item:M2} $\forall \Op{X}_x \in \{\Op{X}_x\}$, the pair $(\ket{\psi}, \{\Op{X}_x\})$ satisfies
\[ \Op{X}_x\ket{\psi} \neq \lambda \ket{\psi},  \quad \lambda \in \mathbb{C}\setminus \{0\}.\]
\end{enumerate}
While finding a suitable pair $(\ket{\psi}, \{\Op{X}_x\})$ is straightforward and the measurement may be simple to perform, the challenge for the verifier lies in ensuring that an untrusted or adversarial prover actually performs the measurement and returns the measurement result. Here, we introduce a randomness certification protocol in which the verifier prepares a single quantum state $\ket{\psi}$, and the prover convinces the verifier they performed measurement $\{\Op{X}_x\}$ on $\ket{\psi}$ by providing a witness -- a random number $x$ with probability distribution $\func{p_{\ket{\psi}}\!}{x}$. We prove a theorem that any verifier can use to certify that $x$ was generated randomly, i.e. obtained from a measurement satisfying \ref{item:M1}-\ref{item:M2}. Importantly the theorem's validity is independent of the prover's physical state. It is genuinely black-box certifiable. Moreover, by removing the need for a quantum computer or entanglement, the protocol is easier, cheaper and less resource intensive than current approaches \cite{Pironio2010, Acin2016, Brakerski2021, Mahadev2022, Aaronson2023, Liu2025}; using just a single particle for each interaction round.

\begin{remark}[A single quantum state]
There are actually two ways one can reduce the randomness of a quantum measurement. First, if contrary to \ref{item:M2}, a measurement is performed where $\Op{X}_x\ket{\psi} = \lambda \ket{\psi}$, then $x$ is observed with unit probability and the outcome of the measurement on this state is not genuinely unpredictable. The second comes about from an ambiguity in how physicists use the term \emph{quantum state}. Say the prover is given $n$ copies of a quantum state $\ket{\psi}$. Then we can either: draw boxes around each $\ket{\psi}$ and say the prover has $n$ quantum states; or we can place the $n$ copies into one box saying the prover has just one quantum state $\ket{\Psi} := \ket{\psi}^{\otimes n}$, where $\otimes$ denotes the tensor (outer) product. Although both descriptions are valid, the statistics differ. A measurement on the state $\ket{\Psi}$ allows many samples to be drawn from the distribution $\func{p_{\ket{\psi}}\!}{x}$, and as $n$ approaches infinity, the prover can perfectly reconstruct $\func{p_{\ket{\psi}}\!}{x}$ allowing them to deterministically return any statistic (e.g. the mean) drawn from this distribution.

In the following, a single quantum state is defined to be a \hyperlink{non-separable}{non-separable} unit vector in a complex Hilbert space $\mathscr{H}$, i.e. any normalised vector that cannot be decomposed into the tensor product of more than one vector. These quantum states are the single-use random boxes of quantum theory.
\end{remark}

\subsection*{Black-box certifiable random number generation}
In any randomized computational model, there are several theoretical and practical differences which demarcate how random numbers are generated. Theoretically, the generation of private randomness is easy, all the verifier need do is use this part of the model. At a practical level, cryptographic security is linked to the ability to generate random numbers locally (private randomness), and does not need to be black-box. True black-box randomness certification in contrast, removes the need to explicitly monitor how the data is produced, allowing trust to be built up in networks of otherwise untrusted parties. This has applications in ensuring the fairness of sampling procedures\footnote{See \url{https://www.americanscientist.org/article/quantum-randomness} for a general introduction.}. More profound perhaps, are the theoretical consequences. Modern classical cryptography and the Natural Proof barrier to the \textsf{P} vs \textsf{NP} problem for instance, assert that \emph{efficiently} distinguishing pseudorandomness from randomness in the black-box setting is impossible\footnote{The emphasis is on \emph{efficient} here. This is because classical cryptography does not use randomness and instead uses \textsf{NP}(-hard) problems with a pseudorandom input. If such problems cannot be efficiently inverted, then \textsf{P} $\neq$ \textsf{NP}.}. Furthermore, treating the prover as a black-box effectively allows them oracle access, hence any proofs relativize to a prover with an oracle.
 
Currently, two main approaches to certifiable randomness exist\footnote{A third protocol requires a full fault-tolerant quantum computer \cite{Brakerski2021, Mahadev2022}.}. In Bell certified randomness (BCR), two spatially separated provers, Alice and Bob, demonstrate a Bell violation \cite{Fine1982, Pironio2010, Hensen2015, Acin2016}. By challenging them to compute a function of two input bits $\func{f}{a,b}$, where each party receives only one bit, the verifier forces them to satisfy \ref{item:M1}-\ref{item:M2} in order to surpass a deterministic bound. In the second approach, circuit sampling randomness (CSR), the verifier sends the prover -- who possesses a quantum computer -- a quantum circuit $\Op{C}$ and asks for the most likely outputs of the circuit (measured in the computational basis) \cite{Aaronson2023, Liu2025}. By imposing time and computational power constraints, together with unproven computational assumptions, the verifier aims to ensure that \ref{item:M1}-\ref{item:M2} were performed in order to sample the circuit's output. However, neither protocol is black-box certifiable. In both cases the verifier must explicitly check the interior workings of the prover in order to certify the data.

In BCR, the verifier must ensure that Alice and Bob cannot communicate (the challenge bit they received), otherwise they can evaluate $\func{f}{a,b}$ deterministically and violate the Bell inequality without randomness. In addition the verifier must also ensure no third party receives the challenge bits, either en-route or transmitted from Alice and Bob, thus allowing communication. In short, the verifier must not only monitor Alice and Bob's location, they must also escort each input bit to Alice and Bob and further ensure nobody modifies their response in order to guarantee randomness. These requirements are not black-box -- a no-signalling constraint is imposed on the prover -- nor can a single, localised verifier certify them.

CSR on the other hand exchanges black-box computability with \emph{feasible} computability. A quantum circuit $\Op{C}$ on $n$ qubits is chosen that is conjectured to be (exponentially) hard to deterministically compute, but computable nevertheless. The issues with this approach are myriad. The verifier assumes, without proof, that $\Op{C}$ cannot be deterministically evaluated in less than $T_c \times \mathrm{flops}/s$ steps and uses this to define a cut-off time before which the prover must respond. However, this assumption is not an absolute bound on $T_c$, rather it is an asymptotic argument which furthermore works against the verifier, since exploiting asymptotic scaling means paying an exponential price in (deterministic) verification. Beyond assuming that sampling from $\Op{C}$ is classically hard, the verifier also assumes a peculiar form of quantum hardness, namely that a quantum computer cannot reliably sample the most likely outputs of $\Op{C}$ \cite{Liu2025}. This second assumption is needed because of the tension introduced by \ref{item:M2}. Observe that in CSR, the most likely output of a circuit can be deterministically computed, meaning strings that score highly on the linear cross entropy benchmark actually have lower entropy. For the prover to convincingly produce a given amount of entropy they must perform a delicate balancing act; ensuring the quantum computer outperforms a classical computer to rule out deterministic function evaluation, but not by too much lest they come up against the determinism implied by \ref{item:M2}. Neither of these statements have been proven in the context of CSR\footnote{An earlier version of this paper presented a loophole claiming the prover can reduce the entropy of their data polynomially in the number of samples from the circuit. As far as I am aware, although the prover can reduce the entropy of their response, it is not currently known how efficiently they can do so and whether such an entropy reduction is possible.}.%Taking the mode is expected to result in a logarithmic entropy reduction with the number of samples. }.

%Imagine replacing $\Op{C}$ with Shor's algorithm to factor a large integer $N$. Now if the prover returns a factor of $N$ before $T_c$ has elapsed the prover may be convinced of \ref{item:M1}, i.e. the factor came from a measurement on a quantum state, but note that this data is not random -- it is a deterministic output of the challenge function. 

But these criticisms distract from the main issue. Granting the assumptions of CSR and pretending that an absolute bound $T_c \times \mathrm{flops}/s \leq W$ for every challenge circuit is proven, the verifier still cannot certify the data as random. The simple reason being that doing so requires checking the prover does not have access to $W$\,flops of compute.
%In the \hyperlink{Appendix}{Appendix}, we show that if these assumptions are indeed correct, then the circuits used in CSR give rise to a class of problems that are computable on a deterministic Turing machine but not on a \hyperlink{quantum Turing machine}{quantum Turing machine}. This could be taken as evidence that quantum computers are less powerful than classical computers. Paradoxically it seems that for randomness certification, the verifier should use a challenge function that is deterministically computable to arbitrary precision but not randomly computable; contradicting the introduction and violating something like the negation of the Church-Turing thesis. The resolution of this apparent paradox is thus. Granting all the assumptions and pretending that an absolute bound $T_c \times \mathrm{flops}/s \leq W$ for every challenge circuit is proven, the verifier still cannot certify the data as random -- the simple reason being that doing so requires checking the prover does not have access to $W$\,flops of compute. %As the following loophole highlights, $W$ is not exponential but polynomial in the input size.

To summarise, because the verifier uses a Turing computable challenge function in both BCR and CSR, the prover's ability to produce a deterministic witness for a challenge function depends on their computational power. If the prover has enough compute, something that cannot be checked in the black-box setting, then they can produce a witness deterministically. In \emph{Projecting Computational Power} \cite{Adleman2018} Leonard Adleman notes that such challenges always query the prover's state, further asking whether any method of avoiding this pitfall exists. These observations motivate some questions:
\begin{enumerate}
\myitem{(Q1)}\label{item:Q1} \textbf{Do truly black-box protocols exist?} Are existing protocols the best one can do in terms of black-box certification?
\myitem{(Q2)}\label{item:Q2} \textbf{Is entanglement necessary?} Known protocols use entanglement as a proxy for quantumness. Is entanglement required for random number certification?
\myitem{(Q3)}\label{item:Q3} \textbf{Is initial randomness required?} Current protocols rely on an initial random seed, they are in fact randomness \emph{expansion} or randomness \emph{amplification} \cite{Vazirani2012, Amer2025}. Does true randomness \emph{generation} exist? 
\end{enumerate}

In the next protocol the prover is asked to estimate something initially in the verifier's possession, meaning the verifier avoids querying the prover's state -- any accepting witness depends only on the verifier's (initial) state. Alternatively, after sending the prover this something, the prover is no longer a black-box. The verifier now `knows' something about the prover and their challenge function should query only this element.

\subsection*{Estimation certified randomness}
\begin{figure}[bth]\hypertarget{Fig1}{}
\includegraphics[width=0.58 \textwidth]{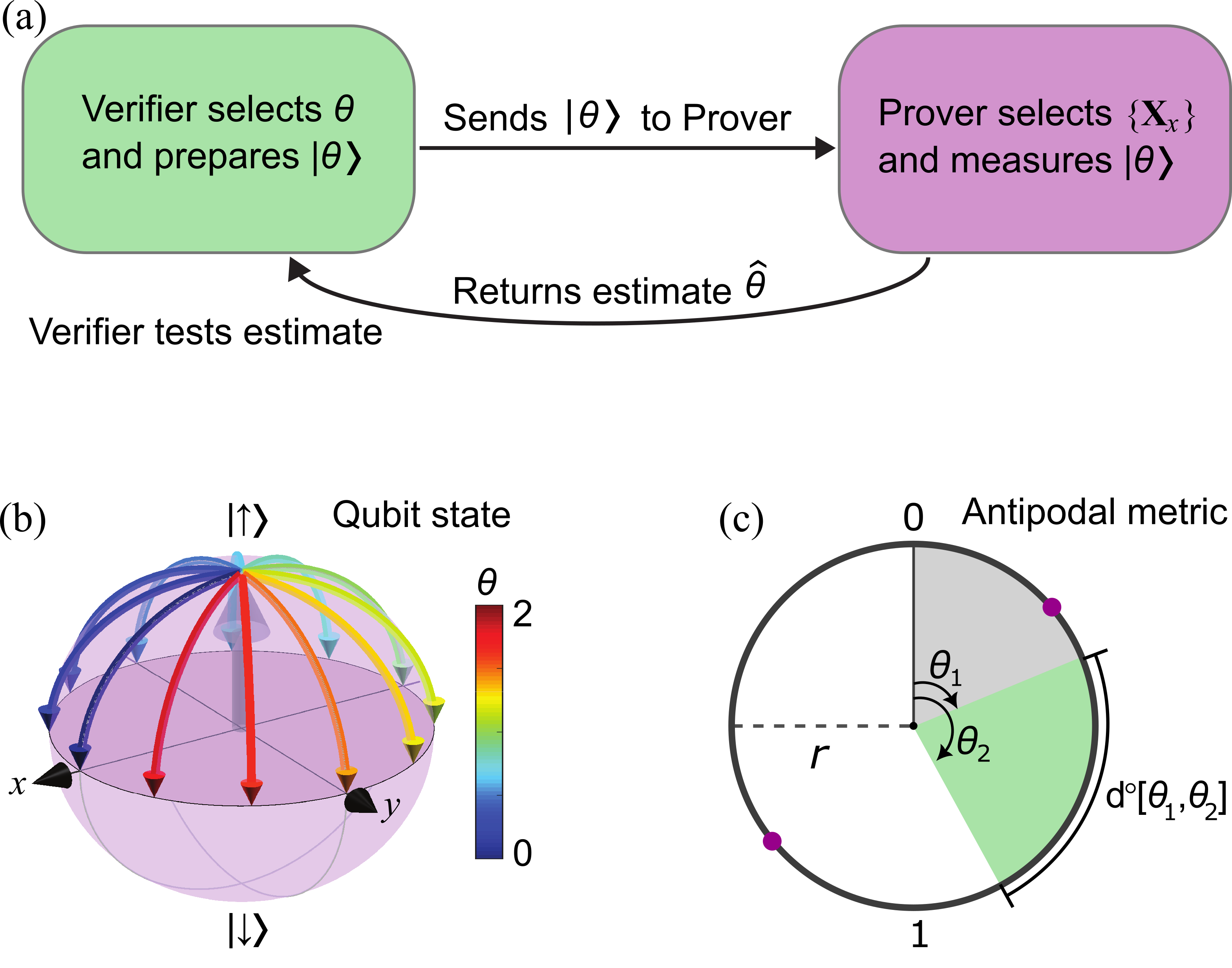}
\caption{\textbf{(Quantum) estimation certified randomness (ECR).} (a) One round of ECR involves: \emph{i)} the verifier choosing a value of $\theta$ and encoding it into a quantum state $\ket{\theta}$, \emph{ii)} sending $\ket{\theta}$ to the prover, \emph{iii)} the prover returning an estimate $\est{\theta}$ of $\theta$ which may or may not derive from a measurement of $\ket{\theta}$, \emph{iv)} the verifier testing the received estimate. (b) A mapping of the parameter $\theta$ to the quantum phase of a qubit, this is a quantum state $\ket{\theta} = \frac{1}{\sqrt{2}}\left(\ket{\uparrow} + \func{exp}{\iu \pi \theta} \ket{\downarrow}\right)$. %This gives a class of states lying on the equator of Bloch sphere, with distance $\antidist{\theta_1,\theta_2}$ between any two states $\ket{\theta_1},\ket{\theta_2}$.  Encoding $\theta$ to a quantum state $\ket{\theta}$. 
(c) Antipodal metric. The antipodal distance $\antidist{\theta_1,\theta_2}$ between two points $\theta_1,\theta_2$ is given by the (shortest) arc length on a circle connecting $\theta_1$ and $\theta_2$. We parametrize the circle by $\theta \in [0, 2)$, the positive angle (with the vertical) in $\pi$ radians. An antipodal pair of points, such as the purple circles, are two points where $\antidist{\theta_1,\theta_2}=\pi r$, and an antipodal probability distribution assigns the same probability to both points of an antipodal pair.}
\label{fig1}
%\end{center}
\end{figure}
In estimation certified randomness (ECR), the verifier prepares a single quantum state $\ket{\theta}$, dependent on a parameter $\theta$ and transmits this state to the prover, asking them to estimate the value of $\theta$ (\hyperlink{Fig1}{Fig.\,1a}). The parameter $\theta$ could, for example, be a quantum phase
\be \label{eq:phase} \ket{\theta} = \frac{1}{\sqrt{2}}\left(\ket{\uparrow} + \func{exp}{\iu \pi \theta} \ket{\downarrow}\right), \ee
between two basis states of some operator $\Op{H}$ in $\mathscr{H}$ (\hyperlink{Fig1}{Fig.\,1b}). In what follows we refer to $\theta$ simply as a phase. If the prover's estimate $\est{\theta}$, of $\theta$ passes a statistical test -- specifically if it's mean squared error is below a certain bound -- then no deterministic protocol could have produced it and the verifier is convinced that $\est{\theta}$ is the result of a probabilistic measurement on $\ket{\theta}$. An example based on classical statistical parameter estimation can be helpful in illuminating the idea behind ECR:
\begin{example}[No-measurement estimation bound for uniform prior and Euclidean error]\label{ex:1}
If the verifier uniformly samples $\theta$ from $[0,2)$ and prepares $\ket{\theta}$ as in \cref{eq:phase}, then without a measurement of $\ket{\theta}$, no estimate has expected mean squared error $\func{\overline{mse}}{\theta,\est{\theta}} < 1/3$. Furthermore, the unique estimate with minimum $\func{\overline{mse}}{\theta,\est{\theta}} = 1/3$, is $\est{\theta}_{\text{opt}} = 1$.
\end{example}
For completeness we include the following definitions:
\begin{definition}[Mean squared error of an estimate]\label{def:mse}
The \emph{mean squared error} (MSE) of an estimate $\est{\theta}$ is defined as
\be \label{eq:mse} \func{mse}{\theta,\est{\theta}} := \evalwrt{\mathrm{p}\!}{\left(\func{d}{\theta,\est{\theta}}\right)^2} = \sum_{\est{\theta} \in \Theta} \func{p}{\est{\theta}} \left(\func{d}{\theta,\est{\theta}}\right)^2, \ee
where $\func{p}{\est{\theta}}$ denotes the (discrete) probability distribution of the estimate, the estimation error -- the difference between $\est{\theta}$ and $\theta$ -- is quantified by a \hyperlink{distance metric}{distance metric} $\func{d}{\theta,\est{\theta}}$, $\Theta$ denotes the set of values that $\theta$, $\est{\theta}$ can take, and we write the expected value with respect to probability distribution $\func{p}{\cdot}$ as $\evalwrt{\mathrm{p}\!}{\cdot}$. Here we write sums for simplicity, one can replace them with integrals for the continuous case.
\end{definition}
\begin{definition}[Expected mean squared error of an estimate]\label{def:emse}
The \emph{expected mean squared error} (EMSE) of an estimate $\est{\theta}$ is defined as
\be \label{eq:emse} \func{\overline{mse}}{\theta,\est{\theta}} := \evalwrt{\mathrm{\pi}\!}{\evalwrt{\mathrm{p}\!}{\left(\func{d}{\theta,\est{\theta}}\right)^2}} = \sum_{\theta \in \Theta} \left(\func{\pi}{\theta} \sum_{\est{\theta} \in \Theta} \func{p}{\est{\theta}} \left(\func{d}{\theta,\est{\theta}}\right)^2\right), \ee
where $\func{\pi}{\theta}$ is the (prior) probability distribution of $\theta$ and in all that follows, the prover selects a real-valued $\theta$ from $[0,2)$\footnote{To avoid confusion, although $\Theta$ should be taken as the set of real numbers (when using Euclidean distance), we allow that $\func{p}{\est{\theta}}$ and $\func{\pi}{\theta}$ be discrete probability distributions meaning they are non-zero on a countable subset of $\Theta$. Later, when using an antipodal metric space, the metric is defined for $\Theta = [0, 2)$.}.
\end{definition}

\cref{ex:1} is a text-book result in statistical estimation, no quantum mechanics is needed and the verifier does not need to prepare $\ket{\theta}$, it is only mentioned for context. The best \emph{a priori} (Euclidean) estimate for the prover is the mean of the prior ($\est{\theta}_{\text{opt}} = \evalwrt{\mathrm{\pi}\!}{\theta}$) with a minimum expected mean squared error equal to the variance of the prior. Without measurement data the prover is constrained, their best estimate is deterministic -- they should always return `1'. %\footnote{Known as the minimizer of the Bayes risk with quadratic cost.}

It is hard to overstate the importance of this elementary result for randomness certification. In fact, with this result we are nearly done. A lower bound on the $\func{\overline{mse}}{\theta,\est{\theta}}$ provides a statistical test the verifier can use. Any estimate with lower expected mean squared error \emph{must} derive from a probabilistic measurement, meaning that if the prover surpasses this (deterministic) bound, the verifier is convinced \ref{item:M1} was performed. To enforce \ref{item:M2} one option for the verifier is to use a prior where every pair $(\ket{\psi}, \{\Op{X}_x\})$ not satisfying \ref{item:M2} has measure zero, for example a continuous uniform prior. %\footnote{A nearly equivalent condition is satisfied by sampling $\Op{C}$ from the Haar measure in CSR \cite{Aaronson2023, Liu2025} although it is a uniform sampling on a discrete set.}. 
We will return to this point later.

The validity of \cref{ex:1} depends on an implicit assumption in statistical parameter estimation that is rarely stated: the prover cannot make use of any information on the value of $\theta$, except for some (probabilistic) measurement data. Because this assumption lies at the centre of any estimation task\footnote{This assumption is not only required for estimation theory, it is baked into computational complexity theory. In BCR and CSR an initial random seed is used, and with prior knowledge of this seed the prover can falsely convince the verifier to certify deterministic data. In fact, whenever a pseudorandom string is used (e.g. in cryptography), \cref{ass:1} is used. The same assumption is being made if we prevent a Turing machine using a look-up table to solve a problem.} and because we are aiming at rigour, we write it explicitly:

\begin{assumption}[No hidden information on $\theta$]\label{ass:1}
Apart from the quantum state $\ket{\theta}$ transmitted by the verifier, the prover has no information on the value of $\theta$. Formally,
\begin{enumerate}
\myitem{(A1)}\label{item:A1} Without $\ket{\theta}$, the prover does not possess any data or function $\func{f}{\cdot}$ dependent on $\theta$.
\myitem{(A2)}\label{item:A2} Given $\ket{\theta}$, the only data (output of a function dependent on $\theta$) that the prover may obtain is the result of a measurement of $\ket{\theta}$ with probability distribution given by \cref{eq:Born}.
\myitem{(A3)}\label{item:A3} (1) and (2) apply to each round of interaction between the prover and verifier\footnote{\ref{item:A3} is not necessary. The verifier does not adapt their choice of $\theta$ based on the response of the prover. They could also choose to send the quantum states for each round of ECR all at once and ask the prover to estimate the vector parameter $\bm{\theta} = \{\theta_1, \theta_2, \cdots , \theta_n\}$. But for didactic purposes it is easier to think of consecutive rounds of interaction.}.
\myitem{(A4)} For completeness, we assume the prover is given the metric space $\MS[\Theta]$ (both the distance function $\dist{\cdot,\cdot}$ and the set of values $\Theta$ it is defined over).% and a description of the prior used to sample $\theta$.
\end{enumerate}
\end{assumption}

In ECR, the verifier samples $\theta$ from a distribution \prior\ and encodes $\theta$ in a quantum state $\ket{\theta}$ before sending it to the prover and receiving an estimate in return; repeating the procedure multiple times with a new $\theta$ each time. The only way the prover's estimate can violate a no-measurement (and deterministic) bound is via a function dependent on $\theta$, namely a measurement of $\ket{\theta}$ with probability distribution $\func{p_{\ket{\theta}}\!}{x}$. With a judicious choice of $\prior$, the verifier can additionally satisfy themselves that \ref{item:M2} holds, i.e. certify the data is random. Moreover, by demanding the best single bit (radix-2) estimate of $\theta$, the verifier can force the prover to use a specific measurement basis consistent with \ref{item:M2} and prevent the prover from sneaking extra determinism into their estimate via the lower significant figures of $\est{\theta}$.

To the reader who remains unconvinced -- perhaps there exists a deterministic way to violate the no-measurement bound? -- we emphasise that such a method would also violate the no-cloning theorem \cite{Wootters1982}, the Heisenberg uncertainty relations \cite{Barnett1989, Aharonov2002}, precision bounds on quantum state tomography \cite{ODonnell2016, Aaronson2018} and the quantum Cram\'er-Rao bound \cite{Helstrom1967, Holevo1982, Braunstein1996}. Together these foundational results collectively assert that, given one copy of an unknown state $\ket{\theta}$, the value of $\theta$ cannot be perfectly and deterministically estimated. Rather, the best estimate $\hat{\theta}$ is derived from a probabilistic measurement, exhibiting non-zero mean squared error and providing randomness.

It turns out, however, that there are technical reasons why these theoretical results are unsuitable for randomness certification. In particular, what it means for a state to be ``unknown'',``arbitrary'' or ``mixed'' is not defined in these works \cite{Wootters1982, Barnett1989, Aharonov2002, ODonnell2016, Aaronson2018, Helstrom1967, Holevo1982, Braunstein1996}, whereas the quantum tomography and quantum Cram\'er-Rao bounds are asymptotic -- they apply in the limit of infinite measurements \cite{ODonnell2016, Aaronson2018, Boixo2007, Jarzyna2015, Pang2017, Gorecki2020}. More problematic is that the quantum Cram\'er-Rao lower bound only applies to unbiased estimates \cite{Cramer1946}, it says nothing about the mean squared error of estimates that systematically under or over-estimate $\theta$. In short, these results are not universally valid, they do not quantitatively address how much randomness can be generated per measurement nor do they relate the estimate mean squared error with the amount of randomness produced.

To overcome these issues, we use a bounded, non-Euclidean metric for ECR, specifically an \hyperlink{antipodal metric}{antipodal metric}. The arc distance between points on a circle, as illustrated in \hyperlink{Fig1}{Fig.\,1c}, is one example of an antipodal metric, and it is isomorphic to the class of states lying on the equator of the Bloch sphere (\hyperlink{Fig1}{Fig.\,1b}). In this metric, the amount of randomness generated correlates with how much the deterministic bound is violated. Another benefit compared to a Euclidean metric, is that larger violations of the no-measurement bound can be observed. This works in the verifier's favour, since they want to make it as easy as possible when deciding if the data is truly random\footnote{A third reason is that the distance between quantum states (as shown in \hyperlink{Fig1}{Fig.\,1b}) is non-Euclidean, and the verifier should use the same metric when assessing the estimate error. I.e. the antipodal metric is the natural metric for differentiating different (quantum) phases.}. 

\begin{definition}[Antipodal (phase) estimation problem]
An estimation problem, where one must estimate the value of an unknown parameter $\theta$, is defined as an \emph{antipodal estimation problem} if:
\begin{enumerate}
\item The distance metric is an \hyperlink{antipodal metric}{antipodal distance metric}, denoted $\antidist{\cdot,\cdot}$.
\item The prior for $\theta$ is an \hyperlink{antipodal probability}{antipodal probability distribution}, denoted $\antiprob{\cdot}$.
\end{enumerate}
If the antipodal metric has a \hyperlink{metric diameter}{metric diameter} of 1, the problem is a \emph{unit antipodal estimation problem}.
%\footnote{Examples of antipodal metrics are the distance between numbers on a clock, directions on a compass, or importantly here the distance between values of $\theta$ in the quantum phase. Whenever the metric has some rotational or periodic symmetry. Antipodal because it is related to antipodal points on a circle, sphere etc. It is important that we use a non-Euclidean metric.}
\end{definition}

Here, as $\theta$ takes on real values in $[0,2)$, the antipodal metric with unit diameter is given by: $\antidist{\theta,\est{\theta}} := \left|\func{sin}{\pi(\theta - \est{\theta})/2}\right|$, and $\antiprob{\theta} = \antiprob{(\theta+1)\,\mathrm{mod}\,2}$, $\forall \theta \in \Theta$. For the unit antipodal estimation problem, the following theorem bounds the expected mean squared error of any string returned by the prover without a measurement of $\ket{\theta}$. 

\begin{genthm}{theorem}{No-measurement bound for antipodal phase estimation}\label{th:anti}
Let $\theta \in [0, 2)$ be sampled from an antipodal probability distribution on a unit antipodal distance metric. Given \cref{ass:1}, then without a measurement of $\ket{\theta}$, no estimate $\est{\theta}$ (deterministic or probabilistic) has expected mean squared error differing from\footnote{Note, \cref{eq:anti} is a lower bound on the expected MSE of the estimate. To see this, exchange “=” with “$\geq$”. The equality here is the special case when the upper bound and lower bound meet.}
\be \label{eq:anti}
\smashfunc{\overline{mse}^\circ}{\theta,\est{\theta}} = \sum_{\theta \in \Theta} \left(\antiprob{\theta} \sum_{\est{\theta} \in \Theta} \func{p}{\est{\theta}} \left(\antidist{\theta,\est{\theta}}\right)^2\right) = 1/2.\ee
Proof: see \hyperlink{Proof1}{Appendix}.
\end{genthm}

The antipodal phase estimation problem has a beautiful property; to generate $\est{\theta}$ the prover can take the output of any deterministic $\func{g}{\cdot} : \mathcal{D} \to \mathbb{R}$ or probabilistic $\func{g}{\cdot} : \mathcal{D} \to (\mathbb{R}, p)$ \emph{estimator} function mapping to the reals. As long as $\func{g}{\cdot}$ does not depend on $\theta$, the output is guaranteed to have $\smashfunc{\overline{mse}^\circ}{\theta,\est{\theta}} = 1/2$. Importantly, \cref{th:anti} does not rely on computational hardness or place constraints on the prover's computational power -- it does not use computability theory. The estimator function $\func{g}{\cdot}$ can be \emph{any} function; it can be \textsf{NP}-complete or non-computable, nor does it need to be enumerable -- the function output can be a real number or \hyperlink{measurable function}{non-measurable}\footnote{Proving this last statement requires the expected value of a set to have one property. If all elements in the set take the same value $a$, then the expected value is $a$.}. The approach here is solely to require $\frac{\partial \func{g}{\cdot}}{\partial \theta} = 0$, for all estimator functions available to the prover.

\subsection*{Demonstration of estimation certified randomness}
We experimentally demonstrate estimation certified randomness in a non-remote setting using the spin state of a single nitrogen-vacancy (NV) center in diamond (\hyperlink{Fig2}{Fig.\,2a}). Here, the verifier selects a phase $\theta$ from an antipodal probability distribution which they map to the NV spin state, using the unit antipodal distance metric to characterise the error of the estimate they receive. Instead of transmitting the state $\ket{\theta}$ to a remote prover, an adjacent prover estimates $\theta$ by measuring the stationary NV spin. Whilst not remote, the protocol is truly black-box, in that the verifier makes no assumptions on the prover, beyond \cref{ass:1}.

\begin{figure}[bth]\hypertarget{Fig2}{}
\includegraphics[width=0.62 \textwidth]{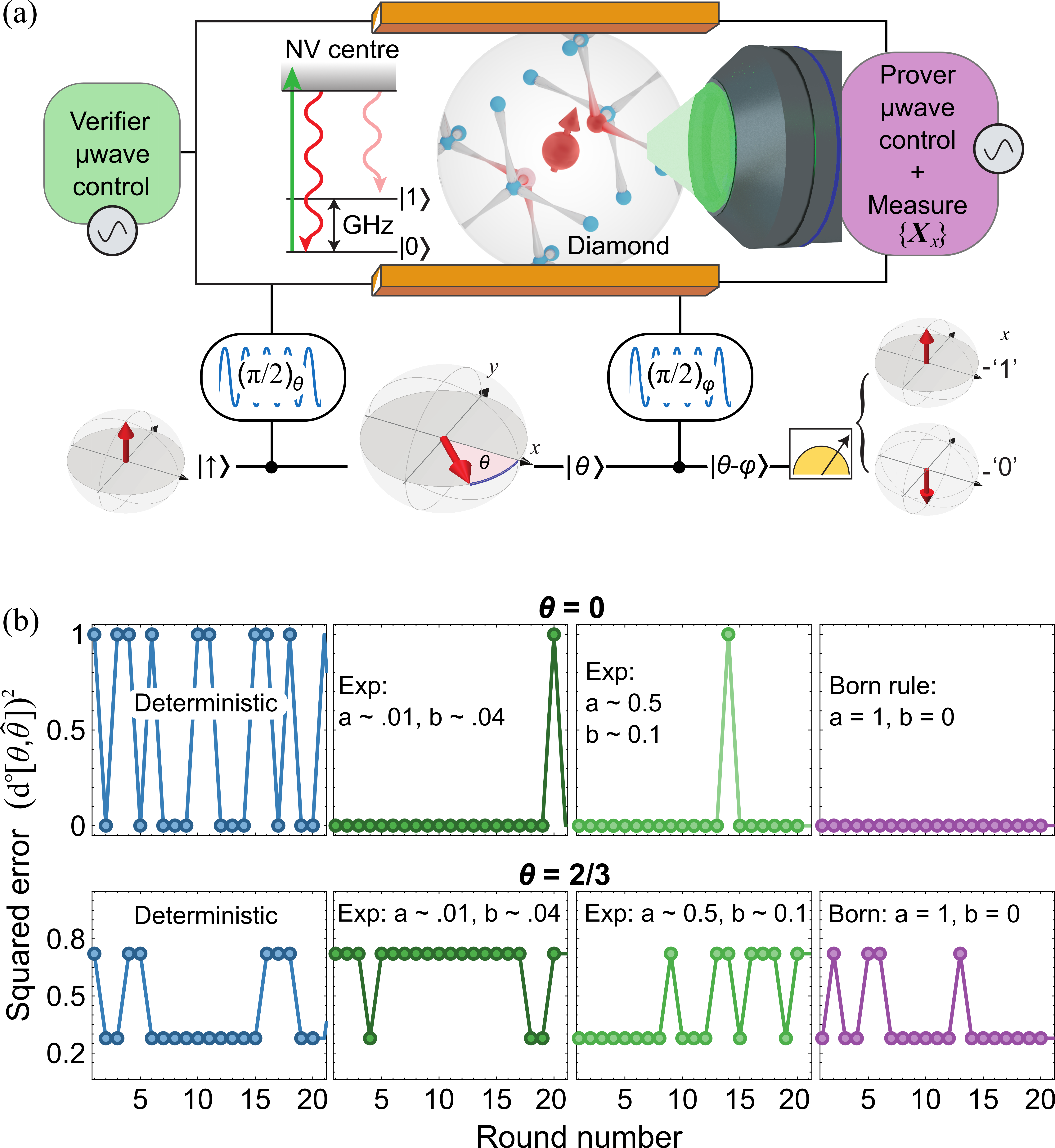}
\caption{\textbf{Demonstration of non-remote estimation certified randomness with a single NV centre in diamond.} (a) After selecting a value of $\theta$, the verifier prepares the NV spin state: $\ket{\theta} = \frac{1}{\sqrt{2}}\left(\ket{\uparrow} + \func{exp}{-\iu \pi \theta} \ket{\downarrow}\right)$ by applying a resonant microwave $\pi/2$-pulse with phase $\pi\times\theta$. The prover performs an $X$-basis measurement of $\ket{\theta}$ by applying a second $\pi/2$-pulse with phase $\pi\times\varphi = 0$ and projectively reading out the spin-state along $z$, denoting the outcomes `1' and `0'. The best estimate of $\theta$ is the measurement outcome. (b) Plots of the individual estimate squared error for 20 rounds of ECR, when the verifier selected $\theta = 0$ (top) and $\theta = 2/3$ (bottom). Four different estimate strategies are shown. A deterministic estimate without a measurement (blue, the sequence of estimates is a permutation of the binary representation of $e^{\pi}$), estimates using the experimental measurement outcome distributed according to \cref{eq:meas_prob} with $a\sim0.01, b\sim0.04$ (dark green) and $a\sim0.5, b\sim0.1$ (light green) and a simulation of the ideal estimate according to the Born rule (purple).}
\label{fig2}
\end{figure}

To map $\theta$ to the NV spin state, the verifier applies a resonant microwave pulse according to the Hamiltonian
\begin{equation}
\Op{H}(t_0,t) = \frac{\hbar \omega_0}{2}\Op{\sigma}_z + \frac{\hbar \Omega}{2} \func{cos}{\omega (t_0 + t) + \phi}\Op{\sigma}_y - \frac{\hbar \Omega}{2}\func{sin}{\omega (t_0 + t) + \phi}\Op{\sigma}_x,
\label{eq:Ham}\end{equation}
where $\Omega$ is the microwave Rabi frequency, $\phi$ the microwave phase at $t_0 = 0$, $\Op{\sigma}_{x,y,z}$ the Pauli operators and $\omega_0$ is the transition frequency between the eigenstates $\ket{\uparrow}$ and $\ket{\downarrow}$ of $\Op{\sigma}_z$ which coincides with the microwave frequency $\omega \sim 1.5$\,GHz. Starting with the spin in $\ket{\uparrow}$, the verifier prepares $\ket{\theta}$ (as in \cref{eq:phase}), by applying $\Op{H}(t_0,t)$ with phase $\phi=(\pi \times \theta)$ from $t_0 = 0$ to $t = \pi/(2\Omega)$ using an arbitrary waveform generator with full phase-control. We emphasise, in the next step, the prover is free to do \emph{anything} they like and in evaluating the estimate, the verifier makes no assumptions on the prover's actions. For exposition we describe the optimal prover strategy (if they choose to co-operate) as performed here.
 
To best estimate $\theta$, the prover uses a second independent signal generator to apply a $\pi/2$-rotation to $\ket{\theta}$ with phase $(\pi\times\varphi)$, resulting in the state
\[\ket{\theta - \varphi} = \Op{U}^{\pi/2}_\varphi \ket{\theta} = \left(\left(1 - e^{\pi \iu (\theta - \varphi)}\right)\ket{\uparrow} + \left(1 + e^{\pi \iu (\theta -\varphi)}\right)\ket{\downarrow}\right)/2,\] followed by projective readout in the $\{\ket{\uparrow}, \ket{\downarrow}\}$ basis (\hyperlink{Fig2}{Fig.\,2a}). Denoting the measurement operators and observables
\[ \{ \Op{Z}_x \}\equiv \left\{\Op{1}_1, \Op{0}_0 \right\}:= \left\{\begin{pmatrix} 1 & 0\\ 0 & 0 \end{pmatrix}, \begin{pmatrix} 0 & 0\\ 0 & 1 \end{pmatrix}\right\}, \quad  \{\Op{\varphi}_x \} := \left\{\Op{1}_1\cdot \Op{U}^{\pi/2}_\varphi, \Op{0}_0\cdot \Op{U}^{\pi/2}_\varphi \right\}, \quad \{x \} = \{1, 0\},\]
the Born rule (\cref{eq:Born}) associates the probability distribution
\be \label{eq:meas_prob} \func{p_{\ket{\theta}}\!}{x = 1} = a\func{sin^2}{\pi(\theta - \varphi)/2} + b, \quad \func{p_{\ket{\theta}}\!}{x = 0} = a\func{cos^2}{\pi(\theta - \varphi)/2}+b,\ee
(with $a = 1, b = 0$) to $x$ when the measurement $\{ \Op{Z}_x \}$ is performed on $\ket{\theta - \varphi}$ (this measurement is denoted $|\Op{\varphi}_x\ket{\theta}|^2$). Restricted to a one-bit estimate, the optimal strategy for the prover is to set $\varphi = 0$, i.e. projective readout of $\ket{\theta}$ in the $X$-basis, and we denote this measurement $\left|\Op{X}_x\ket{\theta}\right|^2$. Experimentally, a confocal microscope is used to optically readout the NV spin state by recording the spin-dependent photoluminescence after 350\,ns green laser excitation \cite{Jelezko2006, Maze2008, Balasubramanian2009} and the prover obtains $\func{p_{\ket{\theta}}\!}{x}$ with $a \sim 0.01, b \sim 0.04$. By swapping the NV electron spin and nuclear spin before reading out the nuclear spin, this is improved to $a \sim 0.52, b \sim 0.09$ \cite{Neumann2010}.

In total, several thousand rounds of ECR were performed, where each round involved the verifier uniformly selecting $\theta$ from $\Theta_6 := \{0, 1, 2, 3, 4, 5\}/3$ and the prover returning a one-bit estimate $\est{\theta}$. In \hyperlink{Fig2}{Fig.\,2b} we show the individual estimate squared errors for a subset of 20 rounds of ECR when the verifier prepared $\ket{\theta}$ with $\theta = 0$ and $2/3$. We compare four different prover strategies. The estimate $\est{\theta} = x$ is the outcome of: \textbf{i)} a deterministic function without a measurement\footnote{The sequence of estimators is a permutation of the binary representation of $e^\pi$.} (blue), \textbf{ii)} a low fidelity measurement $\left|\Op{X}_x \ket{\theta}\right|^2$ (dark green), \textbf{iii)} a higher fidelity measurement $\left|\Op{X}_x \ket{\theta}\right|^2$ (light green), and \textbf{iv)} a simulation of an ideal measurement $\left|\Op{X}_x \ket{\theta}\right|^2$ according to the Born rule (purple). As a low fidelity measurement is equivalent to a low probability to perform a measurement, we can see in the data from this small sample size that the estimate squared error acts as a witness that a quantum measurement was performed.

\hyperlink{Fig3}{Fig.\,3} shows the (posterior) mean squared error of these four estimates strategies for increasing number of ECR rounds. After $n$ rounds of ECR, the verifier calculates the (posterior) mean squared error\footnote{$\theta_i$, $\est{\theta}_i$ are the values of $\theta$ and $\est{\theta}$ in the $i^{th}$ round of ECR.}: $\frac{1}{n} \sum^{n}_{i = 1} \left(\antidist{\theta_i,\est{\theta}_i}\right)^2$ which converges to $\smashfunc{\overline{mse}^\circ}{\theta,\est{\theta}}$ for large $n$; accepting that the data was randomly generated if this value differs from $1/2$ by more than some confidence region. A five-sigma confidence region is shown here, corresponding to a one in several million chance that the verifier falsely accepts a completely deterministic data-set. Experimentally, when the prover used high fidelity readout (\hyperlink{Fig3}{Fig.\,3a} light green), the verifier could be satisfied with five-sigma confidence that the data produced from 120 rounds of ECR was randomly generated, whereas with low fidelity readout, tens of thousands of rounds of ECR were needed before the verifier could reach this confidence (\hyperlink{Fig3}{Fig.\,3b} dark green). Notably, the mean squared error of the deterministic strategy (\hyperlink{Fig3}{Fig.\,3b} blue) approaches the value set by \cref{th:anti} as $n$ increases (as it must), staying within two standard deviations for the entire data-set.

\begin{figure}[bth]\hypertarget{Fig3}{}
%\begin{center}
%\subfigure[]{
\includegraphics[width=0.72 \textwidth]{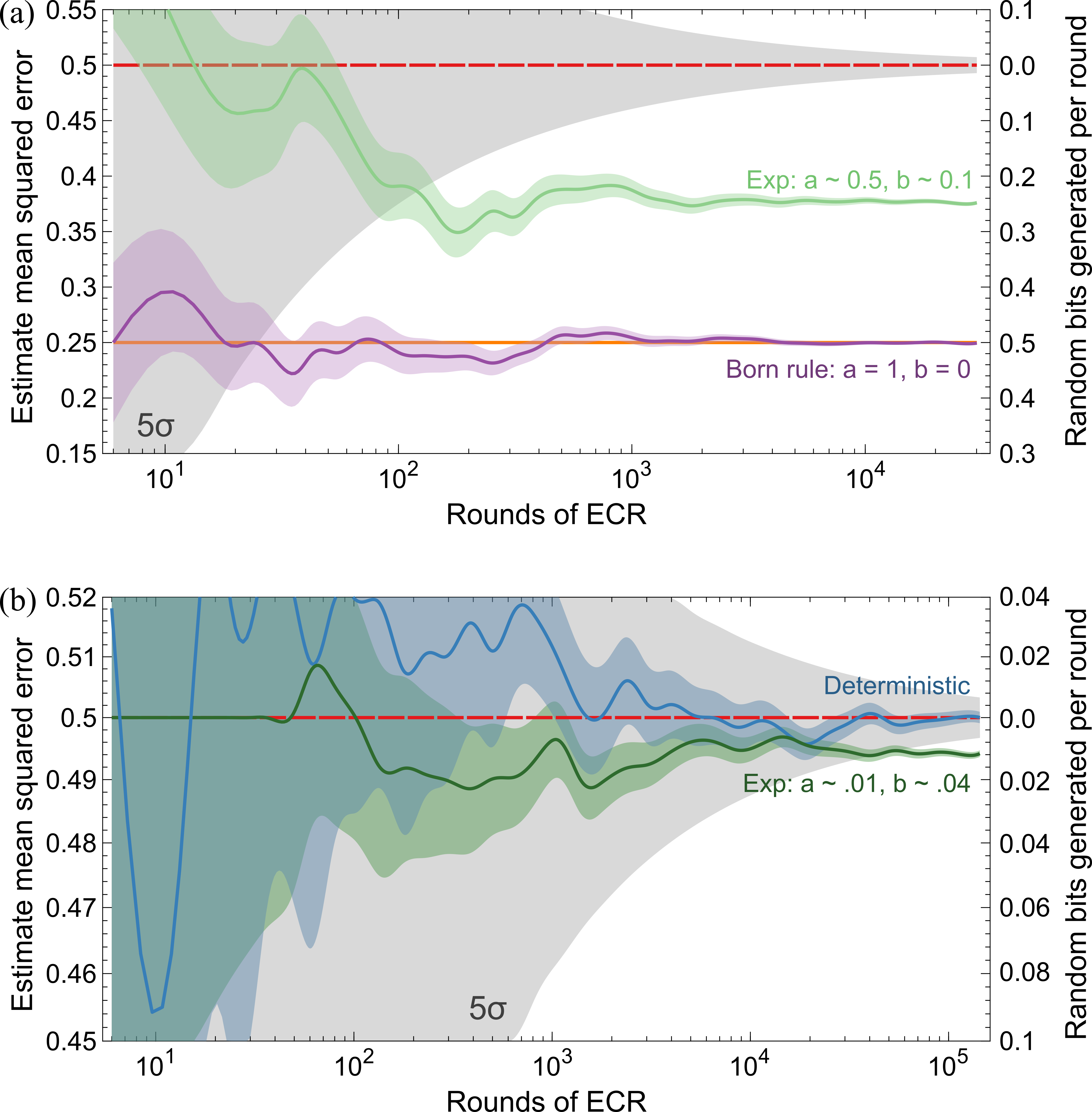}
\caption{\textbf{Experimentally testing the randomness of data.} The estimate mean squared error as a function of the number of ECR rounds when the verifier uniformly selected $\theta$ from $\Theta_6 := \{0, 1, 2, 3, 4, 5\}/3$ and the prover returned a one-bit estimate $\est{\theta}$ based on four different strategies. (a) The prover's estimate is the outcome of a measurement $\left|\Op{X}_x \ket{\theta}\right|^2$ with $a\sim0.5$, $b\sim0.1$ (light green) or a simulation of an ideal measurement with $a = 1$, $b = 0$ (purple). (b) The prover's estimate is the outcome of a measurement $\left|\Op{X}_x \ket{\theta}\right|^2$ with $a\sim0.01$, $b\sim0.04$ (dark green) or a deterministic permutation of the binary representation of $e^\pi$ without a measurement (blue). Shaded regions show one standard deviation region of the data and grey is the five-sigma confidence region for a deterministic estimate to differ from $1/2$.}
\label{fig3}
%\end{center}
\end{figure}

Based on the results of \hyperlink{Fig3}{Fig.\,3} it is natural to ask what estimate minimizes $\smashfunc{\overline{mse}^\circ}{\theta,\est{\theta}}$. I.e. what combination of measurement on $\ket{\theta}$ and statistic of the measurement outcome has minimum expected mean squared error? In the \hyperlink{quantum measurement bound}{Appendix} we prove the following theorem.
\begin{genthm}{theorem}{Quantum measurement bound for antipodal phase estimation}\label{th:CRLB}
Given the quantum state $\ket{\theta} = \frac{1}{\sqrt{2}}\left(\ket{\uparrow} + \func{exp}{\iu \pi \theta} \ket{\downarrow}\right)$ and \cref{ass:1}, the following hold regarding unit antipodal phase estimation:
\begin{enumerate}
\item Over all quantum measurements, the measurement $\left|\Op{\varphi}_x\ket{\theta}\right|^2$ on $\ket{\theta}$ is optimal. No other measurement on $\ket{\theta}$ admits an estimate with lower expected mean squared error.
\item If $\theta$ is sampled uniformly from $[0, 2)$, the estimate $\est{\theta} = (\varphi + x) \,\text{mod}\,2$ obtained from $\left|\Op{\varphi}_x\ket{\theta}\right|^2$ has minimum $\smashfunc{\overline{mse}^\circ}{\theta,\est{\theta}} = 1/4$. In particular, no measurement on $\ket{\theta}$ produces data, and no estimate of $\theta$ based on this data has lower expected mean squared error. Restricting to radix-2 estimates $\est{\theta} \in \{0, 1\}$, further imposes $\varphi \in \{0, 1\}$ as the unique set of measurements achieving this minimum (with $\est{\theta} = x$ when $\varphi = 0$).
\item If $\theta$ is uniformly selected from an $N>2$ point set such that $\sum^N_{k=1}e^{2 \pi \iu \theta_k} = 0$ with $N$ even, then the best estimate derived from a measurement on $\ket{\theta}$ has $\smashfunc{\overline{mse}^\circ}{\theta,\est{\theta}} = 1/4$, achieved with $\est{\theta} = (\varphi + x) \,\text{mod}\,2$. %, for all $\varphi$ (i.e. independent of the readout phase). And t
The best radix-2 estimate satisfies this with $\est{\theta} = x$ when $\varphi = 0$.
\end{enumerate}
\end{genthm}
The results of \hyperlink{Fig3}{Fig.\,3} (purple line) correspond to \cref{th:CRLB}--(3) with $N = 6$, proving that this is the optimal co-operative strategy; if the prover could realize an ideal measurement. \cref{th:CRLB} gives a Heisenberg-type uncertainty relation regarding how much randomness ECR can certifiably generate per round. One measurement on a single quantum state cannot both: \emph{i)} produce uniformly random data and \emph{ii)} produce data correlated with $\theta$. Unable to simultaneously return two mutually incompatible or non-commuting observables -- prove a measurement of $\ket{\theta}$ was performed \emph{and} prove randomness -- then, on average, each round of ECR can yield half a certifiably random bit, at most\footnote{Perfectly satisfying \emph{i)} leads to $\smashfunc{\overline{mse}^\circ}{\theta,\est{\theta}} = 1/2$, which cannot be verified, whereas perfectly satisfying \emph{ii)} gives $\smashfunc{\overline{mse}^\circ}{\theta,\est{\theta}} = 0$, implying exact knowledge of $\theta$ and no randomness.}. \cref{th:CRLB} also allows the verifier to quantify (actually bound) the entropy content of the data they receive. Given a string $x$, the amount of \emph{verifiable entropy} in $x$ is maximised by minimizing its expected mean squared error. Although strings with higher entropy exist, for example $x$ drawn uniformly from $\{0,1\}$, the verifier cannot distinguish them from deterministic strings. At the other extreme, further improving the expected mean squared error to below the bound of \cref{th:CRLB} would lower the string's entropy, but quantum mechanics forbids such an estimate.

\subsection*{Discussion}
We have demonstrated a protocol that answers questions \ref{item:Q1} and \ref{item:Q2} in that it is truly black-box certifiable and does not require entanglement. We now address \ref{item:Q3}, whether initial randomness is required. Again, \cref{ass:1} does all of the heavy lifting, removing the need for randomness or pseudorandomness from the verifier. The verifier is free to deterministically choose any value of $\theta \in \Theta$ (where $\prior \neq 0$) and \cref{ass:1} guarantees the prover does not know the value chosen regardless of the method used to make the selection. The proofs of \cref{th:anti} and \cref{th:CRLB} are valid for any deterministic procedure that makes $n$ selections from $\Theta$ as long as for all $\theta_i$ the ratio $k_i \theta_i/n = \antiprob{\theta_i}$, where $k_i$ is the number of times $\theta_i$ is selected\footnote{Note that the realisation of such a procedure requires a discrete prior probability distribution.}. In fact, in \hyperlink{Fig3}{Fig.\,3} for each round of ECR, the verifier deterministically selected $\theta_i$ sequentially from $\Theta_6$, so that the mean squared error converged smoothly as a function of $n$. One could argue that most discussions of pseudorandomness have it backwards, it is a mechanism used to (hopefully) ensure that \cref{ass:1} holds.

Although some might be uneasy with \cref{ass:1}, its rejection is akin to giving the prover abilities above and beyond anything else considered here. For example, \cref{ass:1} exposes the flawed logic in the introduction. Yes, if the prover has access to the infinite set of all deterministic Turing machines, this will include machines that output a string to pass any randomized test. But at a meta-level the prover still needs to make a choice of which Turing machine to use. If this choice itself is made by a deterministic Turing machine, we are left with asking how the selection is performed? \Cref{ass:1} places a restriction on the selection procedure, without invoking it not only can the prover return the solution to any problem, they can do so without ever being given the problem instance; for example returning the sum of two integers without receiving those integers (from the verifier) as an input.

\subsection*{Summary}
It is widely believed that the Church-Turing thesis is true and that genuine black-box random number certification is impossible. We have introduced a generalized function that both allows for certification of randomness from a black-box and contradicts the Church-Turing thesis; the problem considered here cannot be evaluated on a deterministic Turing machine to the same error as on a (quantum) randomized Turing machine. Although it risks being treated as a tautology
\begin{quotation}
\emph{Q: What can't a deterministic Turing do? A: Produce a random output.}
\end{quotation}
the implications are significant.

% Ties to my 1TF conjecture, what can entanglement not do. 

%With a probabilistic random variable, there is a function of $\theta$ that returns lower error than any deterministic function. 

In proving randomness generation, we arrived at a stronger parameter estimation bound than current methods. Compared to the Cram\'er-Rao lower bound which is universally employed in quantum metrology \cite{Helstrom1967, Holevo1982, Braunstein1996, Boixo2007, Jarzyna2015, Pang2017, Gorecki2020} the bound derived here can be saturated with a single measurement and does not assume an unbiased estimator. Under the same conditions (a single measurement on a single non-separable quantum state) it is known that Shor's algorithm must yield a mean squared error for phase estimation that improves exponentially in time and number of qubits \cite{Kitaev2002, Nielsen2000, Preskill2015, Atia2017}. In contradiction the quantum Cram\'er-Rao lower bound forbids such a precision, it is linear in time and number of qubits \cite{Helstrom1967, Holevo1982, Braunstein1996, Boixo2007, Jarzyna2015, Pang2017, Gorecki2020}. Resolving this contradiction in favour of the stronger estimation bounds would exponentially reduce the power and usefulness of quantum entanglement for computation.

Finally, the results here hold promise of a new approach to attacking the \textsf{P} vs \textsf{NP} problem. No diagonalization argument was needed and the expected mean squared error was bounded without the Cauchy-Schwarz inequality. We introduced two techniques: \emph{i)} we restrict the prover's initial information on a solution by forbidding \emph{a priori} access to any function dependent on the solution, additional information is only obtained by running an algorithm which terminates in a measurement, and \emph{ii)} we use the circular phase symmetry inherent in quantum states. In doing so we could bound absolutely (not asymptotically) the precision of any deterministic or probabilistic machine for evaluating certain classes of problems. This phase symmetry is not unique to quantum mechanics, by Noether's theorem -- conservation of probability implies a phase symmetry -- it is present in any probabilistic model of computation.

\bibliographystyle{naturemag}
\bibliography{references2025.bib}%{}

\newpage

\section*{\hypertarget{Appendix}{Appendix}}
\subsection*{No-measurement bound on antipodal phase \hypertarget{Proof1}{estimation}}
\cref{th:anti}: Given that \cref{ass:1} holds. Let $\theta \in [0, 2)$ be sampled from an \hyperlink{antipodal probability}{antipodal probability distribution}: $\forall \theta \in \Theta$, $\antiprob{\theta} = \antiprob{(\theta+1)\,\mathrm{mod}\,2}$ with \hyperlink{antipodal metric}{antipodal distance metric} of diameter 1: $\antidist{\theta,\est{\theta}} := \left|\func{sin}{\pi(\theta - \est{\theta})/2}\right|$. Then without $\ket{\theta}$, no estimate $\est{\theta}$ (deterministic or probabilistic) has expected mean squared error less than
\[ \smashfunc{\overline{mse}^\circ}{\theta,\est{\theta}} = \sum_{\theta \in \Theta} \left(\antiprob{\theta} \sum_{\est{\theta} \in \Theta} \func{p}{\est{\theta}} \left(\antidist{\theta,\est{\theta}}\right)^2\right) = 1/2.\]
That is, any estimate such that $\func{p}{\est{\theta}}$ does not depend on $\theta$, has $\smashfunc{\overline{mse}^\circ}{\theta,\est{\theta}} = 1/2$.

\begin{proof}[Proof of \cref{th:anti}] Here assuming, $\antiprob{\theta}$ has discrete probability mass function (the proof follows analogously for a continuous probability distribution, simply exchange the summation $\sum_{\theta \in \Theta}$ with integration $\int_{\Theta}\,\mathrm{d}\theta$):
%Given that $\theta \in [0, 2)$ is sampled from an antipodal probability distribution with antipodal distance metric, then without $\ket{\theta}$, no estimate $\est{\theta}$ (deterministic or probabilistic) has expected mean squared error less than
%\be \label{eq:anti}
%\smashfunc{\overline{mse}^\circ}{\theta,\est{\theta}} = \sum_{\theta \in \Theta} \left(\antiprob{\theta} \sum_{\est{\theta} \in \Theta} \func{p}{\est{\theta}} \antidist{\theta,\est{\theta}}^2\right) = 1/2.\ee
\begin{align*} \label{eq:anti}
\smashfunc{\overline{mse}^\circ}{\theta,\est{\theta}} &= \sum_{\theta \in [0,2)} \left(\antiprob{\theta} \sum_{\est{\theta} \in \Theta} \func{p}{\est{\theta}} \left(\antidist{\theta,\est{\theta}}\right)^2\right)\\
& = \sum_{\theta \in [0,1)} \left(\antiprob{\theta} \sum_{\est{\theta} \in \Theta} \func{p}{\est{\theta}} \left(\grayub{\func{sin^2}{\pi(\theta - \est{\theta})/2}}_{\theta \in [0, 1)} + \grayub{\func{sin^2}{\pi(\theta + 1 - \est{\theta})/2}}_{\theta \in [1, 2)}\right)\right)\\
& = \sum_{\theta \in [0,1)} \left(\antiprob{\theta} \sum_{\est{\theta} \in \Theta} \func{p}{\est{\theta}} \left(\grayub{\func{sin^2}{\pi(\theta - \est{\theta})/2} + \func{cos^2}{\pi(\theta - \est{\theta})/2}}_{= 1}\right)\right)\\
& = \sum_{\theta \in [0,1)} \left(\antiprob{\theta} \grayub{\sum_{\est{\theta} \in \Theta} \func{p}{\est{\theta}}}_{=1}\right) = \sum_{\theta \in [0,1)} \antiprob{\theta} = 1/2, 
\end{align*}
where the second and last equalities make use of the fact that $\antiprob{\theta}$ is an antipodal probability distribution.
\end{proof}
Note, we can prove \cref{th:anti} without needing to restrict the estimator to $\est{\theta} \in [0, 2)$. I.e. \cref{th:anti} holds for any estimate $\est{\theta} \in \mathbb{R}$ (as long as we define $\left|\func{sin}{\pi(\theta - \est{\theta})/2}\right|$ over $\mathbb{R}$, although strictly speaking this distance function is a \hyperlink{distance metric}{pseudo-metric} over the reals). The proof can be generalized for an antipodal phase estimation problem with an arbitrary antipodal (pseudo-)metric, not necessarily on the reals, with \hyperlink{metric diameter}{metric diameter} $d$. If $\func{p}{\est{\theta}}$ does not depend on $\theta$, then we have:
\[ \smashfunc{\overline{mse}^\circ}{\theta,\est{\theta}} = d^2/2.\]
The proof proceeds by pairing every point $x$ in $(\Theta, \mathrm{d}^\circ)$ with it's antipodal pair $y$, observing that $\antiprob{\theta}$ is the same for both points of an antipodal pair, and substituting $d^2 = \left(\antidist{x,\est{\theta}}\right)^2 + \left(\antidist{y,\est{\theta}}\right)^2$ for any $\est{\theta} \in \Theta$ when $x, y$ are an antipodal pair.

%\begin{assumption}[The estimate and the parameter are on the same metric space]
%Additional assumption is that $\theta$ and $\est{\theta}$ are elements in a metric space (can even be a pseudo-metric space), and their probability measures are defined on the metric space if they probabilistic. The verifier can enforce this by requiring that $\est{\theta} \in \{0,1\}$, or any subset of the metric space. For any estimate that is not in the metric space the verifier can assign emse of $1/2$ to these estimates or reject these estimates.
%\end{assumption}

\subsection*{\hypertarget{quantum measurement bound}{Quantum measurement} for antipodal phase estimation}
Obtaining a universal estimation bound that applies to any estimator derived from any measurement is actually a difficult proposition. Especially if we would like the bound to be achievable. This section breaks down the analysis needed to prove \cref{th:CRLB} into two parts. First we look at the family of probability distributions that a measurement on the quantum state $\ket{\theta}$ can produce. Second we consider how one can transform the measurement outcome (i.e. the probability distribution) to give the best estimate of $\theta$. This is the class of classical (post-processing) functions on the measurement data, these functions are sometimes called estimators.

\subsubsection*{The family of probability distributions produced by a quantum measurement on $\ket{\theta}$}
	As $\ket{\theta}$ is a quantum state in a 2-dimensional Hilbert space, the outcome $x$ of a \emph{projective} measurement on $\ket{\theta}$ satisfies
\[ \func{p_{\ket{\theta}}\!}{x = 0} = |\braket{\Phi,\varphi}{\theta}|^2, \quad \quad \func{p_{\ket{\theta}}\!}{x = 1} = 1- |\braket{\Phi,\varphi}{\theta}|^2,\]
where $\bra{\Phi,\varphi}$ is an arbitrary quantum state in the (dual) Hilbert space $\mathscr{H}$ and we denote the measurement outcomes as 0 and 1. Note, the parameters $\Phi$ and $\varphi$ are selected by the prover when performing the measurement. The random variable $x$, resulting from such a measurement has a probability distribution with dependence on $\theta$ of the general form 
\be \label{eq:pdf} \func{p_{\ket{\theta}}\!}{x = 0} = \Phi\func{cos}{\pi (\theta-\varphi)}+1/2 \quad \text{and} \quad \func{p_{\ket{\theta}}\!}{x = 1} = 1/2 - \Phi\func{cos}{\pi(\theta-\varphi)},\ee% which we can write succinctly as
%\be\label{eq:opt_pmf} \func{p}{x; \theta} = \func{cos^2}{\frac{\pi}{2}(\theta-\varphi + x)},\quad \forall x \in \{0, 1\}\ee
with $0 \leq \Phi \leq 1/2$.

We now change our notation to align it with the notation that is common in statistical parameter estimation; using \px\ to denote a probability distribution that depends on $\theta$. By \cref{ass:1} the only dependence on $\theta$ comes from a measurement on $\ket{\theta}$, so we have $\px \equiv \func{p_{\ket{\theta}}\!}{x}$. In fact, by using $\theta$ to parametrize the probability distribution, as we have done in \cref{eq:pdf}, we can constrain the form of \emph{any} general quantum measurement in a Hilbert space of arbitrary dimension, not just a projective measurement\footnote{These more general measurements are called positive operator valued measurements (POVM).}. I.e any quantum measurement that produces additional measurement outcomes, not just a two outcome distribution. As only $\ket{\theta}$ depends on $\theta$, any other measurement outcomes must have a uniform distribution. Let $Y$ denote the set of these outcomes, then every element $y \in Y$ has a constant probability distribution: $\forall \theta$, $\func{p}{y; \theta} = b$, $\forall y\in Y$,  with a respective reduction of $\Phi$ in \cref{eq:pdf}. This observation is closely connected to results showing the Fisher information on $\theta$ is maximised for a projective measurement and cannot be increased with a general POVM. The two outcome measurement described by \cref{eq:pdf} with $\Phi = 1/2$ maximises the Fisher information on $\theta$ (c.f. \cref{eq:meas_prob} with $a = 1, b=0$), no other measurement can result in a probability distribution with greater Fisher information. This is one reason the verifier should restrict the prover to a single bit estimate in ECR, when sending a two dimensional quantum state $\ket{\theta}$. Any additional bits of information do not derive from a quantum measurement on $\ket{\theta}$.

\subsubsection*{The best estimate using the measurement outcome}
We now consider the set of functions the prover can apply to the measurement outcome in order to best estimate the value of $\theta$. There is actually a blurred line distinguishing whether the prover's actions take effect before the measurement is performed, thus altering the measurement probability distribution, or they take effect after and can be considered post-processing of the measurement outcome. Much of this analysis could be (and in effect was) covered in the previous section.

Observe that for the antipodal phase estimation problem with measurement outcomes distributed according to \px, if any estimate exists with $\smashfunc{\overline{mse}^\circ}{\theta,\est{\theta}} < 1/2$, then randomly throwing away the measurement data only increases the expected mean squared error. The reasoning is straight-forward, if the prover throws away data, then the estimate that would have derived from this data must be replaced with one that has $\smashfunc{\overline{mse}^\circ}{\theta,\est{\theta}}= 1/2$, increasing the expected mean squared error. We now formally state and prove the proposition.

\begin{genthm}{proposition}{High-fidelity measurement for antipodal estimation -- the best estimate does not throw away measurement data}\label{prop:fidelity}
Given \cref{ass:1} and the unit antipodal estimation problem, let $x\in \mathcal{B}_x$ denote the outcome of a measurement on the state $\ket{\theta}$ with \hyperlink{probability space}{probability measure} \px\ on \hyperlink{measurable space}{measurable space} \MB. If \px\ admits any estimate with $\emse{\theta,\est{\theta}} < 1/2$ then a reduction of the probability measure \px\ by including results from a disjoint measurable space increases the expected mean squared error of this estimate.
\end{genthm}

\begin{proof}
When data is thrown away these elements must be replaced by something else. Let $\func{p}{y}$ be an arbitrary probability measure on a disjoint \hyperlink{measurable space}{measurable space} $(Y, \mathcal{B}_y)$, such that $Y \cap X = \varnothing$. By \cref{ass:1}, $\func{p}{y}$ cannot depend on $\theta$. Write the expected mean squared error of an estimate (allowing both the use of measurement data and any other probabilistic or deterministic function) as a convex combination of these two probability measures \px\ and $\func{p}{y}$\footnote{Any finite convex combination of probability measures is another probability measure.}:

\begin{align*} \smashfunc{\overline{mse}^\circ}{\theta,\est{\theta}} &= \alpha \times \left(\int_{\Theta} \antiprob{\theta}  \sum_{x\in X} \px \left(\antidist{\theta,\est{\theta}}\right)^2 \,\mathrm{d}\theta \right) + (1- \alpha) \times \grayub{\left( \int_{\Theta} \antiprob{\theta} \sum_{y\in Y} \func{p}{y} \left(\antidist{\theta,\est{\theta}}\right)^2 \,\mathrm{d}\theta \right)}_{=1/2, \;\; \text{\cref{th:anti}}} \\
&= \alpha \times \left( \int_{\Theta} \antiprob{\theta} \sum_{x\in X} \px \left(\dist{\theta,\est{\theta}}\right)^2 \,\mathrm{d}\theta \right) + \frac{1}{2}(1- \alpha).
\end{align*}
Assume there exists a function $\func{g}{\cdot} : X \to \Theta$ such that
\[ \int_{\Theta} \antiprob{\theta} \sum_{x\in X} \px \left(\dist{\theta,\func{g}{x}}\right)^2 \,\mathrm{d}\theta < 1/2.\]
Then the expected mean squared error for an estimate $\est{\theta} = \func{g}{\cdot}$ is always minimised when $\alpha = 1$.
\end{proof}
We proved the high-fidelity proposition for the case that the prior has a continuous probability distribution and the antipodal metric has unit diameter. The proposition also holds for a discrete prior and when the measurement outcomes have a continuous distribution (exchange relevant integrals and sums), and also for an antipodal metric of any diameter $d$ (set $1/2 \to d^2/2$).

One might be led to conclude that the optimal estimate must have a probability distribution $\func{p}{\est{\theta}} = \px$, i.e. the best estimate uses all of the measurement data and does not throw away any information. However, the above only excludes uniformly throwing away data, we need to rule out the possibility that cherry-picking -- selectively throwing away data -- cannot improve the expected mean squared error. Formally: %(have the issue that the null set has probability measure zero. I.e. we cannot ever observe the null data-set if we are using a probability measure...)
\begin{genthm}{proposition}{No cherry-picking for antipodal estimation -- the best estimate does not selectively throw away measurement data}\label{prop:cherry}
%If any estimate has EMSE $<1/2$, then the best estimate uses all the measurement data
Given \cref{ass:1} and the unit antipodal phase estimation problem with measurement outcomes distributed according to probability measure $\func{p}{x_i;\theta}$. There always exists a mapping\footnote{This is a mapping to another measurable space $(\Theta, \mathcal{B}_{\est{\theta}})$ with the same probability measure $\func{p}{\cdot}$, giving the \hyperlink{probability space}{probability space} $(\Theta,\mathcal{B}_{\est{\theta}}, \func{p}{\cdot}$).} $\func{g}{\cdot}: X \to \Theta$ such that for all measurement outcomes $x_i \in X$,
\[\int_{\Theta} \antiprob{\theta} \func{p}{x_i;\theta} \left(\antidist{\theta,\func{g}{x_i}}\right)^2 \,\mathrm{d}\theta \leq 1/2.\]
As a result, the optimal estimate must have probability measure $\func{p}{x_i;\theta}$. More correctly, no other probability measure (not dependent on $\theta$) admits an estimate with lower expected mean square error.
\end{genthm}

\begin{proof}
Let $0 < \func{p}{x_1,\theta} < 1$ denote the probability of observing the measurement outcome $x_1$. %, with conditional/marginal/posterior probability
%\[ \func{p}{x_1 | \theta} = \int_\Theta \antiprob{\theta} \func{p}{x_1;\theta} \,\mathrm{d}\theta \equiv \beta.\]
Partition $X$, the set of measurement outcomes, to write the expected mean squared error as
\begin{align*} \smashfunc{\overline{mse}^\circ}{\theta,\est{\theta}} &= \left(\int_{\Theta} \antiprob{\theta}  \sum_{x\in X\setminus x_1} \px \left(\antidist{\theta,\est{\theta}}\right)^2 \,\mathrm{d}\theta \right) + \alpha \times \left(\int_{\Theta} \antiprob{\theta} \func{p}{x_1;\theta} \left(\antidist{\theta,\func{g}{x_1}}\right)^2 \,\mathrm{d}\theta \right)\\ &+ (1- \alpha) \times \grayub{\left( \int_{\Theta} \antiprob{\theta} \sum_{y\in Y} \func{p}{y} \left(\antidist{\theta,\est{\theta}}\right)^2 \,\mathrm{d}\theta \right)}_{=1/2, \;\; \text{\cref{th:anti}}},
\end{align*}
where we allow $x_1$ to be replaced with elements from $Y$ with probability measure $\func{p}{y}$. If there exists a function $\func{g}{\cdot}$ that maps $x_1$ to a real number $\func{g}{x_1} \in \Theta$ such that
\[  \int_{\Theta} \antiprob{\theta} \func{p}{x_1,\theta} \left(\antidist{\theta,\func{g}{x_1}}\right)^2 \,\mathrm{d}\theta \leq 1/2,\]
then the expected mean squared error is always minimised when $\alpha = 1$, meaning this outcome is never thrown away (for the edge case of equality $= 1/2$, $\alpha$ can be any value and the best estimate can use this data or throw it away). If the above holds for all $x_i \in X$, then the optimum estimate never throws away any of the data, i.e. the optimum statistic does not post-select on the measurement results.

We now show that, for any probability function $\func{p}{x_1,\theta}$ there always exists a real number $\func{g}{x_1} \mapsto r \in \Theta$ with this property. To see this, assume by contradiction, that for all $r \in \Theta$
\[  \int_{\Theta} \antiprob{\theta} \funcwrt{p}{x_1\!}{r,\theta} \left(\antidist{\theta,r}\right)^2 \,\mathrm{d}\theta > 1/2.\]
Take an \hyperlink{antipodal metric}{antipodal pair} $r, s$, then
\begin{align*} \int_{\Theta} \antiprob{\theta} \funcwrt{p}{x_1\!}{r,\theta} \left(\antidist{\theta,r}\right)^2 \,\mathrm{d}\theta + \int_{\Theta} \antiprob{\theta} \funcwrt{p}{x_1\!}{s,\theta} \grayub{\left(\antidist{\theta,s}\right)^2}_{= 1 -\left(\antidist{\theta,r}\right)^2} \,\mathrm{d}\theta &>1\\
\implies \int_{\Theta} \antiprob{\theta} \left(\funcwrt{p}{x_1\!}{r,\theta} \left(\antidist{\theta,r}\right)^2 + \funcwrt{p}{x_1\!}{r,\theta}\left(1 - \left(\antidist{\theta,r}\right)^2\right)\right) \,\mathrm{d}\theta &>1\\
\implies \int_{\Theta} \antiprob{\theta} \funcwrt{p}{x_1\!}{r,\theta} \,\mathrm{d}\theta &>1.
\end{align*}
The last expression is an integral over two probability distributions, which cannot be greater than 1, and we have a contradiction. As $x_1$ was arbitrary, the above holds for any measurement outcome.
\end{proof}
The reason this proof technique works is because for any number $r$ with squared distance to $\theta$ greater than $1/2$, the antipode to $r$ has squared distance less than $1/2$ to $\theta$. I expect we can strengthen this proposition to show that if the measurement outcome has \emph{any} dependence on $\theta$, then there exists a mapping with expected mean squared error $< 1/2$ and not just $\leq 1/2$. Meaning it is impossible to find a probability distribution with some dependence on $\theta$ (is not uniform on all of $\Theta$), where every real number $r$ has $\smashfunc{\overline{mse}^\circ}{\theta,r}= 1/2$. We are now ready to prove \cref{th:CRLB}.

\subsubsection*{Proof of \cref{th:CRLB}}
\Cref{th:CRLB}: Given the quantum state $\ket{\theta} = \frac{1}{\sqrt{2}}\left(\ket{\uparrow} + \func{exp}{\iu \pi \theta} \ket{\downarrow}\right)$ and \cref{ass:1}, the following hold regarding unit antipodal phase estimation:
\begin{enumerate}
\item Over all quantum measurements, the measurement $\left|\Op{\varphi}_x\ket{\theta}\right|^2$ on $\ket{\theta}$ with probability distribution \cref{eq:meas_prob} with $a=1, b = 0$ (c.f. \cref{eq:pdf} with $\Phi = 1/2$) is optimal. No other quantum measurement on $\ket{\theta}$ admits an estimate with lower expected mean squared error.% the random variable $x$ resulting from this measurement has maximum Fisher information over all quantum measurements on $\ket{\theta}$.
%\item Restricted to a radix-2 estimate $\est{\theta} \in \{0, 1\}$, the measurement $\left|\Op{X}_x\ket{\theta}\right|^2$ is the unique optimal measurement. I.e. $\left|\Op{\varphi}_x\ket{\theta}\right|^2$ with $\varphi \in \{0, 1\}$.
\item If $\theta$ is sampled uniformly from $[0, 2)$, the estimate $\est{\theta} = (\varphi + x) \,\text{mod}\,2$ obtained from $\left|\Op{\varphi}_x\ket{\theta}\right|^2$ has minimum $\smashfunc{\overline{mse}^\circ}{\theta,\est{\theta}} = 1/4$. In particular, no measurement on $\ket{\theta}$ produces data, and no estimate of $\theta$ based on this data has lower expected mean squared error. Restricting to radix-2 estimates $\est{\theta} \in \{0, 1\}$, further imposes $\varphi \in \{0, 1\}$ as the unique set of measurements achieving this minimum (with $\est{\theta} = x$ when $\varphi = 0$).
\item If $\theta$ is uniformly selected from an $N>2$ point set such that $\sum^N_{k=1}e^{2 \pi \iu \theta_k} = 0$ with $N$ even, then the best estimate $\est{\theta} = (\varphi + x) \,\text{mod}\,2$ has $\smashfunc{\overline{mse}^\circ}{\theta,\est{\theta}} = 1/4$. %, for all $\varphi$ (i.e. independent of the readout phase). And t
The best radix-2 estimate satisfies this with $\est{\theta} = x$ when $\varphi = 0$.
\end{enumerate}

\begin{proof}[Proof of \cref{th:CRLB}--(1)]
Set $\Phi = 1/2$ in \cref{eq:pdf}, which we can write succinctly as
\be\label{eq:opt_pmf} \func{p}{x; \theta} = \func{cos^2}{\frac{\pi}{2}(\theta-\varphi + x)},\quad \forall x \in \{0, 1\}.\ee
A value of $\Phi < 1/2$ reduces this probability measure and is equivalent to randomly throwing away this measurement outcome, \cref{prop:cherry} (and \cref{prop:fidelity}) guarantee that this will not result in an estimate with lower expected mean squared error, so \cref{eq:opt_pmf} is optimal. No other probability distribution obtained from a quantum measurement of $\ket{\theta}$ admits an estimate with lower expected mean squared error.
\end{proof}
%if \cref{eq:opt_pmf} admits any estimate with $\smashfunc{\overline{mse}^\circ}{\theta,\est{\theta}} < 1/2$, then \cref{prop:fidelity} and \cref{prop:cherry} guarantee that 

The next two parts of \cref{th:CRLB} involve finding a function $\est{\theta} = \func{g}{\cdot}$ (where $\func{g}{\cdot}$ does not depend on $\theta$) that minimizes $\smashfunc{\overline{mse}^\circ}{\theta,\est{\theta}}$ 
%\[ \smashfunc{\overline{mse}^\circ}{\theta,\est{\theta}} = \evalwrt{\pi}{\evalwrt{\mathrm{p}\!}{\left(\antidist{\theta,\est{\theta}}\right)^2}},\]
for the given antipodal prior probability distribution and with $\func{p}{x; \theta}$ given by \cref{eq:opt_pmf}. Writing $\est{\theta} = \func{g}{x}$ and substituting \cref{eq:opt_pmf}
\[ \smashfunc{\overline{mse}^\circ}{\theta,\func{g}{x}} = \int_\Theta \antiprob{\theta} \evalwrt{\mathrm{p}\!}{\left(\antidist{\theta,\func{g}{x}}\right)^2}\,\mathrm{d}\theta = \int_\Theta \func{\pi}{\theta} \sum_{x=0,1}\func{cos^2}{\frac{\pi}{2}(\theta-\varphi + x)}\left(\antidist{\theta,\func{g}{x}}\right)^2\,\mathrm{d}\theta.\]
With antipodal distance function $\antidist{\theta,\est{\theta}} = \left|\func{sin}{\frac{\pi}{2}(\theta-\est{\theta})}\right|$ we have
\begin{align} \label{eq:emse2} 
&\smashfunc{\overline{mse}^\circ}{\theta,\func{g}{x}} = \int^2_0 \antiprob{\theta} \sum_{x=0,1}\func{cos^2}{\frac{\pi}{2}(\theta-\varphi + x)}\left(\func{sin^2}{\frac{\pi}{2}(\theta-\func{g}{x})}\right)\,\mathrm{d}\theta \\ \nonumber
&= \int^2_0 \antiprob{\theta} \left(\grayub{\func{cos^2}{\frac{\pi}{2}(\theta-\varphi + 1)}}_{\func{p}{x=1;\theta}} \func{sin^2}{\frac{\pi}{2}(\theta-\func{g}{1})} + \grayub{\func{cos^2}{\frac{\pi}{2}(\theta-\varphi)}}_{(1-\func{p}{x = 1;\theta})} \func{sin^2}{\frac{\pi}{2}(\theta-\func{g}{0})} \right)\mathrm{d}\theta.%\\
%&= \int^2_0 \antiprob{\theta} \left(\func{sin^2}{\frac{\pi}{2}(\theta-\varphi)}\func{sin^2}{\frac{\pi}{2}(\theta-\func{g}{1})} + \func{cos^2}{\frac{\pi}{2}(\theta-\varphi)}\func{sin^2}{\frac{\pi}{2}(\theta-\func{g}{0})} \right)\mathrm{d}\theta.
\end{align}
%Let $\func{g}{x} = (\varphi+x)\,\mathrm{mod}2$
%\begin{align*}
%\smashfunc{\overline{mse}^\circ}{\theta,\func{g}{x}} &= \int^2_0 \antiprob{\theta} \left(\func{sin^2}{\frac{\pi}{2}(\theta-\varphi)}\func{cos^2}{\frac{\pi}{2}(\theta-\varphi)} + \func{cos^2}{\frac{\pi}{2}(\theta-\varphi)}\func{sin^2}{\frac{\pi}{2}(\theta-\varphi)} \right)\mathrm{d}\theta\\
%&= 4 \int^1_0 \antiprob{\theta} \func{sin^2}{\frac{\pi}{2}(\theta-\varphi)}\func{cos^2}{\frac{\pi}{2}(\theta-\varphi)}\mathrm{d}\theta.%\\
%&= \frac{4}{\pi}\int^\pi_0 \func{\mathrm{\pi}}{\varphi+2u/\pi}\func{cos^2}{u} \func{sin^2}{u} \,\mathrm{d}u.
%\end{align*}
%For any $\varphi$ such that the above integral is greater than $1/2$, the antipode to $\varphi$ yields an integral less than $1/2$.

\begin{proof}[Proof of \cref{th:CRLB}--(2)]
Let $\antiprob{\theta} = 1/2$, $\forall \theta \in [0,2)$ in \cref{eq:emse2}, then
\begin{align*}
\smashfunc{\overline{mse}^\circ}{\theta,\func{g}{x}}& = \frac{1}{2}+\frac{1}{2}\int^2_0  \func{sin^2}{\frac{\pi}{2}(\theta-\varphi)}\left(\func{sin^2}{\frac{\pi}{2}(\theta-\func{g}{1})} - \func{sin^2}{\frac{\pi}{2}(\theta-\func{g}{0})} \right)\mathrm{d}\theta \\
& = \frac{1}{2} + \frac{1}{8}\left(\func{cos}{\pi(\func{g}{1}-\varphi)} - \func{cos}{\pi(\func{g}{0}-\varphi)}\right)
\end{align*}
which attains its minimum of $1/4$ when $\func{g}{1} - \varphi = 1$ and $\func{g}{0} - \varphi = 0$. By \cref{th:CRLB}--(1), no other measurement probability distribution admits an estimate with lower expected mean squared error, so $\smashfunc{\overline{mse}^\circ_{\mathrm{min}}}{\theta,\est{\theta}} = 1/4$ and the estimate $\est{\theta} = (\varphi+x)\,\mathrm{mod}2$ achieves this minimum.

If the prover is restricted to a 2-radix estimate, i.e. $\func{g}{\cdot}: X \to \{0, 1\}$, then the expected mean squared error attains a minimum of $1/4$ when $\varphi = 0$ and $\func{g}{1} = 1$, $\func{g}{0} = 0$ (also when $\varphi = 1$ $\implies$ $\func{g}{1} = 0$, $\func{g}{0} = 1$).
\end{proof}

\begin{proof}[Proof of \cref{th:CRLB}--(3)]
Let $\theta$ uniformly take on values from an $N=2M$ point set (with $M>1$), i.e. $\{0 + \epsilon, 1/M +\epsilon, \dotsc, (2M-1)/M + \epsilon\} \equiv \Theta_{N}$, with $\epsilon \in [0,M)$. Denote $\theta_k = \epsilon + k/M$ for $k = 0, 1, \dotsc, N-1$ and set $\frac{\pi}{2}(\theta_k -\varphi) = \frac{\pi}{2}(\epsilon - \varphi) + \frac{\pi k}{N}$, then \cref{eq:emse2} becomes
\begin{align*}% \label{eq:6phases}
\smashfunc{\overline{mse}^\circ}{\theta,\func{g}{x}} & = \frac{1}{2}+\sum_{k = 0}^{N-1} \frac{1}{N} \func{sin^2}{\frac{\pi}{2}(\epsilon - \varphi) + \frac{\pi k}{N}}\func{sin^2}{\frac{\pi}{2}(\epsilon - \func{g}{1}) + \frac{\pi k}{N}}\\
&- \sum_{k = 0}^{N-1} \frac{1}{N} \func{sin^2}{\frac{\pi}{2}(\epsilon - \varphi) + \frac{\pi k}{N}}\func{sin^2}{\frac{\pi}{2}(\epsilon - \func{g}{0}) + \frac{\pi k}{N}}\\
& = \frac{1}{2} + \frac{1}{8}\func{cos}{\pi \left(\func{g}{0}-\varphi\right)}-\frac{1}{8}\func{cos}{\pi \left(\func{g}{1}-\varphi\right)}.
\end{align*}
The expected mean squared error is minimized when $\func{g}{0} = 0 + \varphi$ and $\func{g}{1} = 1 + \varphi$, i.e. for an estimate $\est{\theta} = x+ \varphi$, with minimum mean squared error: $\smashfunc{\overline{mse}^\circ_{\mathrm{min}}}{\theta,\est{\theta}} = 1/4$.

Restricted to the best 2-radix estimate, then $\smashfunc{\overline{mse}^\circ}{\theta,\est{\theta}} = 1/4$ when $\varphi = 0$ (setting $\est{\theta} = x$) or when $\varphi = 1$ (setting $\est{\theta} = x+1\,\text{mod}\,2$).
\end{proof}
For odd $N$, if $\theta$ is sampled uniformly from an $N>2$ point set such that $\sum^N_{k=1}e^{2 \pi \iu \theta_k} = 0$, and the measurement data is given by \cref{eq:opt_pmf}, then the best estimate $\est{\theta} = x + \varphi$ has $\smashfunc{\overline{mse}^\circ}{\theta,\est{\theta}} = 1/4$, for all $\varphi$ (i.e. independent of the readout phase). However, this is not an antipodal probability distribution and we have only shown that \cref{eq:opt_pmf} is optimal when the prior is an antipodal probability distribution, hence the condition that $N$ be even.

%\newpage
\subsection*{Glossary}
%\textbf{A function is defined as a mapping from a fixed domain to a fixed co-domain. A mapping to a different domain or co-domain (even if it is identical where the domains agree) is a different function. We make this explicit by often writing the domain of a function in the function name, as a superscript before the name appears.} It is a subtle point, but sometimes it can be important or critical.\\
%There is an important issue which is rarely addressed, sometimes the domain or co-domain of a function is written much broader than it truly is. For example we say that a function maps to the reals, when it only maps to the non-negative reals. In computer science if we are told that actual true domain of a function, this could sometimes allow us to compute the function much faster than if we are not told this information. Often we are given a rule and shown how it operates on a small domain and the co-domain is explicitly defined, then we are expected to inductively/recursively increase the rule when the domain is extended, but now we are not told the co-domain, part of the task is to find the co-domain. Makes it a partial function.
%
%When defining the domain and co-domain of a function, the elements of the sets are assumed to have already been defined. When the sets $X$, $Y$ etc are presented, their elements are assumed to have already been defined. Then the powersets (and their elements) and the $\sigma$-algebras (and their elements) are defined.
\begin{definition}[\hypertarget{Quantum Randomized Turing Machine}{Quantum Randomized Turing Machine}]
We define a \emph{quantum randomized Turing machine} as an extension of a deterministic Turing machine, one in which a source of randomness is added. A quantum randomized Turing machine is a deterministic Turing machine that can additionally generate quantum states, send quantum states to another quantum randomized Turing machine in an interactive protocol and perform quantum measurements on quantum states. The outcome of a measurement on a quantum state is a random variable according to the \hyperlink{Born rule}{Born rule}.
\end{definition}

The computational power of a quantum randomized Turing machine is not fully defined here. For example, it is open as to whether the measurement output can be used to solve arbitrary functions, possibly even the Halting problem? Additional components are required to characterize the power of this machine, for instance that transformations on quantum states are unitary and obey the Schr\"odinger equation and that the energy required to perform these transformations and generate quantum states be less than some finite constant. Part of the subtlety in defining the power of a computational model enters when restricting the class of allowable transformations. It is common to allow these higher-level set of (control) transformations to imbue the machine with some of it's computational power. One standard definition of a %\hyperlink{quantum Turing machine}{quantum Turing machine} 
quantum Turing machine for example, allows for a complex transformation matrix with entries such that the real and imaginary parts can be computed to within $2^{-n}$ in time polynomial in $n$ \cite{Bernstein1997}. We leave this part of the definition open to allow the prover access to arbitrary computational power, all that is needed for the results of the paper is that the quantum randomized Turing machine is at least as powerful a deterministic Turing machine, which is achieved in the above definition. The key property that we make use of in \cref{ass:1} is to ensure these transformations do not depend on $\theta$.% \textcolor{blue}{The rotations that allow perfect estimation of $\theta$ in this paper can be computed in polynomial time, thus from the strict definition of a quantum Turing machine, it is not explicitly forbidden to achieve perfect estimation of $\theta$. That is the difference of our model. Quite common in prover-verifier interactions, for example zero-knowledge proofs, despite the name allow polynomial prior knowledge, not exactly zero knowledge.}

Note that the set of quantum states and measurements needed here is small, we are not hiding any computational complexity in the continuum of a complex Hilbert space. The demonstration of estimation certified randomness (ECR) requires that the verifier to generate quantum states from a discrete set of six elements (and the results hold for $\theta \in \{0, 1/2, 1, 3/2\}$, a set of cardinality 4). Additionally, the prover need only perform a two outcome quantum measurement and we do not assume either the preparation of the quantum state or the measurement is perfect. Furthermore, the results remain valid if we allow that the prover can perform an arbitrary quantum measurement to arbitrary precision, this is an element in an uncountable class of sets, related to $SU(2)$.

\begin{remark}[Distinction from probabilistic Turing machines]
A probabilistic Turing machine is generally defined as a deterministic Turing machine augmented with the random outcome of a coin toss. Some models allow that the coin toss be biased, whereas other models, Sipser for example \cite{Sipser2012}, restrict the probability to be $1/2$. I believe that the reason why certifiable black-box random number generation hasn't been realised in the model of a probabilistic Turing machine is because the randomness component of this model is under-defined. The definition is somewhat incomplete. It should be possible to bias the coin toss, but the definition doesn't address how to do this, whether a biased coin can be sent to another deterministic Turing machine and how the bias can be determined. There is no discussion if the randomness component can be sent between interacting parties, or how the probability distribution can be tuned. The axioms of quantum mechanics can be thought of a set of logical rules that answer these questions in a consistent framework. In this context they should not bec considered a physical limitation on a Turing machine, but as a set of rules that govern probabilistic computers.
\end{remark}

\begin{definition}[\hypertarget{Born rule}{Born rule}]
Given a quantum state $\ket{\psi}$ in a complex Hilbert space $\mathscr{H}$, the probability to measure an outcome $x$ associated with a Hermitian operator $\Op{X}_x$ is
\[ \func{p_{\ket{\psi}}\!}{x} := \bra{\psi}\Op{X}_x^{\dagger}\Op{X}_x\ket{\psi}.\]
Here assuming discrete measurement outcomes $x \in X$ in the (countable) \hyperlink{probability space}{probability space} \PS, described by a countable collection $\{ \Op{X}_x\}$ of Hermitian operators that satisfy the completeness relation $\sum_{x\in X}\Op{X}_x^{\dagger}\Op{X}_x = \Op{I}$, where each $\Op{X}_x$ is an element of the space of linear (Hermitian) operators on $\mathscr{H}$
\[ \mathcal{L}(\mathscr{H}) = \{ \Op{X}_x : \mathscr{H} \to \mathscr{H} \text{ such that }\Op{X}_x \text{ is linear}\}.\]
It is standard in quantum theory to associate the measurement outcome $x$ with the real eigenvalue of the operator $\Op{X}$. Here we allow general measurement outcomes.
\end{definition}

\begin{definition}[\hypertarget{non-separable}{Non-separable} quantum state]
A quantum state $\ket{\Psi}$, a unit length vector in a complex Hilbert space $\mathscr{H}$, is defined as a \emph{non-separable} quantum state if and only if it cannot be decomposed into the (outer) tensor product of two quantum states, each in Hilbert spaces of dimension greater than 1 (and lower dimension than $\mathscr{H}$). Conversely, if
\[ \ket{\Psi} = \ket{\psi_1} \otimes \ket{\psi_2}, \text{ where } \ket{\psi_1} \in \mathscr{H}_1 \text{ with } \func{dim}{\mathscr{H}_1} > 1, \;  \text{ and } \ket{\psi_2} \in \mathscr{H}_2 \text{ with } \func{dim}{\mathscr{H}_2} > 1,\]
then $\ket{\Psi}$ is a \emph{separable} quantum state.
\end{definition}

\begin{remark}[Non-separable states]
There is a clear connection between non-separable quantum states and prime numbers -- they both cannot be factored into smaller units. This suggests that we treat non-separable quantum states as the atomic or indivisible units in our model. Note also that just as the trivial decomposition of a prime number by a factor of 1 does not make the number composite, the tensor product of a non-separable quantum state with a unit dimensional Hilbert space does not produce a separable state vector. For example, in optical interferometry quantum states are often written in the occupation number basis $\ket{n}$, with $n$ denoting the number of particles in a given mode. Sometimes the empty or vacuum state of another mode is considered and such a state would be written $\ket{n}\otimes\ket{0}$. If we only consider the vacuum state and never add any photons to this mode, then the Hilbert space of this mode has unit dimension and can be disregarded allowing us to write $\ket{n}$ without loss of generality.
\end{remark}

\begin{definition}[\hypertarget{distance metric}{Distance metric} and a metric space]
Given a set of points $M$, a \emph{distance metric} $\dist{\cdot,\cdot}: M \times M \to [0, \infty]$, defined for all $M$, is a function that operates on two elements of $M$ returning a value in $\extnneg$, that satisfies the following axioms.

For all points $x, y, z \in M$:
\begin{enumerate}
\myitem{(D1)}\label{item:D1} (Identity of indiscernibles) $\dist{x,x} = 0$.
\myitem{(D2)}\label{item:D2} (Positivity) If $x \neq y$, then $\dist{x,y} > 0$.
\myitem{(D3)}\label{item:D3} (Symmetry) $\dist{x,y} = \dist{y,x}$.
\myitem{(D4)}\label{item:D4} (Triangle inequality) $\dist{x,z} \leq \dist{x,y} + \dist{y,z}$.
\end{enumerate}

The pair \MS\ is called a \emph{metric space}. Allowing $\func{d}{x,y} = +\infty$, we have an extended metric, yielding an \emph{extended metric space}. A metric space where \ref{item:D2} is relaxed to allow $\dist{x,y} = 0$ for $x\neq y$ is called a \emph{pseudo-metric space}.
\end{definition}

\begin{definition}[\hypertarget{metric diameter}{Metric diameter}]
The \emph{metric diameter} $\norm{\Md}$ of a metric space \MS\, is defined as the supremum (in the extended reals) of the set of all distances between points in $M$. I.e 
\[ \norm{\Md} := \funclims{sup}{\extnneg}{\{\dist{x,y} : x,y \in M \}}.\]
The assumption that $\dist{\cdot,\cdot}$ is defined on all pairs in $M$ and maps to $\extnneg$ implies that the set of all distances is a subset of the extended reals. Assuming completeness of the extended reals we are guaranteed that $\norm{\Md}$ exists and is well-defined.
\end{definition}

\begin{definition}[\hypertarget{antipodal metric}{Antipodal metric} space and \hypertarget{antipode point}{antipode} points]\label{def:antipodal}
A point $x$ in a metric space \MS\ is defined as an \emph{antipode point} if and only if there exists a point $y \in \MS$ such that the following holds
\[ \left(\dist{x,y}\right)^2 = \left(\dist{x,z}\right)^2 + \left(\dist{y,z}\right)^2,\; \forall z \in \MS. \]
By definition, $y$ is also an antipode point, and the pair $\{x,y\}$ are defined as an \emph{antipodal pair}. A metric space \MS, is defined as \emph{antipodal}, if and only if every point $x \in \MS$ is an antipode point.
\end{definition}

\begin{remark}[Antipodal metric spaces]
One property of antipodal metric spaces that can come in useful is that every point $x \in M$ is a diameter point. See Neumaier \cite{Neumaier1981} for more remarks.
\end{remark}

\begin{definition}[\hypertarget{antipodal probability}{Antipodal probability} distribution (on a metric space)]
Let \MS\ be an antipodal metric space, a probability distribution $\func{p}{x}$, either discrete or continuous, defined for all points in \MS\ is defined as an \emph{antipodal probability distribution} if and only if it assigns the same probability to both points of an antipodal pair. I.e. let $\{x,y\}$ be an antipodal pair, then $\func{p}{\cdot}$ is an antipodal probability distribution if and only if $\func{p}{x} = \func{p}{y}$ for all antipodal pairs.
\end{definition}

The following measure related definitions are adapted from Tao \cite{Tao2011}.

\begin{definition}[\hypertarget{measurable space}{Measurable space}, \hypertarget{measurable set}{measurable set} and \hypertarget{sigma algebra}{$\sigma$-algebra}]
Let $X$ be a set. A \emph{$\sigma$-algebra} on $X$ is a collection $\mathcal{B}$ of subsets of $X$ which obeys the following properties:
\begin{enumerate}
\item (Empty set) $\varnothing \in \mathcal{B}$.
\item (Complement) If $E \in \mathcal{B}$, then the complement $\comp{E} := X\setminus E$ is also in $\mathcal{B}$.
\item (Countable unions) If a countable collection of sets $E_1, E_2, \dotsc \in \mathcal{B}$, then their union is also in $\mathcal{B}$, i.e. $\cup^\infty_{n = 1} E_n \in \mathcal{B}$.
\end{enumerate}
The pair $(X,\mathcal{B})$ is a \emph{measurable space} and elements of $\mathcal{B}$ are called \emph{measurable sets}. A measurable space with respect to a specific algebra $\mathcal{B}$ is called \emph{$\mathcal{B}$-measurable} and often the Borel $\sigma$-algebra is assumed. We write a $\sigma$-algebra on $X$ as $\mathcal{B}_x$ to emphasize that it is defined on subsets of $X$.
\end{definition}

%\begin{definition}[\hypertarget{Borel measurable}{Borel measurable} and \hypertarget{Borel $\sigma$-algebra}{Borel sigma algebra}]
%Let $\Gamma$ be a topological space (possibly a metric space). The \emph{Borel $\sigma$-algebra} $\func{\mathcal{B}}{\Gamma}$ of $\Gamma$ is defined to be the $\sigma$-algebra generated by the open subsets of $\Gamma$. Elements of $\func{\mathcal{B}}{\Gamma}$ are called \emph{Borel measurable}.
%\end{definition}

\begin{definition}[A \hypertarget{measure}{measure}, a \hypertarget{measure space}{measure space} and a \hypertarget{probability space}{probability space}]
Let \MB\ be a measurable space. An (unsigned countably additive) \emph{measure} $\func{\mu}{\cdot}$ is a map $\func{\mu}{\cdot}:\mathcal{B}_x\to [0, +\infty]$ that obeys the following axioms:
\begin{enumerate}
\item (Empty set) $\func{\mu}{\varnothing} = 0$.
\item (Countable disjoint additivity) Whenever $E_1, E_2, \dotsc \in \mathcal{B}_x$ are a countable collection of disjoint measurable sets, then $\func{\mu}{\cup^\infty_{n=1} E_n} = \sum^\infty_{n=1} \func{\mu}{E_n}$.
\end{enumerate}
A triplet $(X,\mathcal{B}_x,\func{\mu}{\cdot})$, where \MB\ is a measurable space and $\func{\mu}{\cdot}$ is an (unsigned countably additive) measure, is called a \emph{measure space}. If the measure has the additional property:
\begin{enumerate}
\setcounter{enumi}{2}
\item (Unit measure) $\func{\mu}{X\setminus\varnothing} = 1 = \func{\mu}{X}$.
\end{enumerate}
I.e. the measure of the (universal) set $X$ is one, then $\func{\mu}{\cdot}$ is a \emph{probability measure}, often written $\func{p}{\cdot}$, and the triplet \PS\ is a \emph{probability space}.
\end{definition}

\begin{definition}[(Borel) \hypertarget{measurable function}{measurable function}]
Let \MB\ be a (Borel) \hyperlink{measurable space}{measurable space} and let $\func{f}{\cdot}:X \to \extnneg$ be an unsigned function. Denote the (Borel) \hyperlink{sigma algebra}{$\sigma$-algebra} on the non-negative extended reals as $\mathcal{B}_{0,\infty}$. Then $\func{f}{\cdot}$ is (Borel) \emph{measurable} if the co-domain of $\funcup{f}{-1}{\cdot}$ is the $\sigma$-algebra $\mathcal{B}_x$. That is, for every open (sub)set $U \in \mathcal{B}_{0,\infty}$ of the non-negative extended reals, the inverse image $\funcup{f}{-1}{U}$ maps to an element in $\mathcal{B}_x$ (is $\mathcal{B}_x$-measurable).

Here the inverse image $\funcup{f}{-1}{S}$ on a set $S$ is the set of all elements in the domain of $\func{f}{\cdot}$ that map to $S$ in the co-domain of $\func{f}{\cdot}$.
\end{definition}

\end{document}